\documentclass[11pt]{amsart}
\usepackage{geometry}                
\usepackage{subfig}
\usepackage{graphicx}
\usepackage{amssymb,mathrsfs,bbold,array}
\usepackage{epstopdf}
\usepackage{multirow}
\usepackage[foot]{amsaddr}

\theoremstyle{remark}

\newcommand{\bo}[1]{\boldsymbol{#1}}
\DeclareMathOperator{\diag}{diag}

\title{Using the Geometric Phase to Optimise Planar Somersaults}
\author{William Tong$^{1,2}$ and Holger R.~Dullin$^1$}
\address{${}^1$School of Mathematics and Statistics, The University of Sydney\newline
$\phantom{*}\;\;\;{}^2$Australian Institute of Health Innovation, Macquarie University}
\email{william.tong@mq.edu.au, holger.dullin@sydney.edu.au}

\begin{document}

\begin{abstract}
We derive the equations of motion for the planar somersault, which consist of two additive terms. The first is the dynamic phase that is proportional to the angular momentum, and the second is the geometric phase that is independent of angular momentum and depends solely on the details of the shape change.
Next, we import digitised footage of an elite athlete performing 3.5 forward somersaults off the 3m springboard, and use the data to validate our model. We show that reversing and reordering certain sections of the digitised dive can maximise the geometric phase without affecting the dynamic phase, thereby increasing the overall rotation achieved. Finally, we propose a theoretical planar somersault consisting of four shape changing states, where the optimisation lies in finding the shape change strategy that maximises the overall rotation of the dive. This is achieved by balancing the rotational contributions from the dynamic and geometric phases, in which we show the geometric phase plays a small but important role in the optimisation process.
\end{abstract}

\maketitle


\section{Introduction}
The somersault is an acrobatic manoeuvre that is essential in Olympic sports such as diving, trampolining, gymnastics and aerial skiing. Today, athletes seek to better understand the scientific theory behind somersaults in the hope of gaining an edge in competition. Within diving alone there is extensive literature ranging from books aimed at athletes and coaches (e.g.\! \cite{batterman, Fairbanks, obrien}), to journal articles targeting the scientific community. Yeadon has provided great insight into the biomechanics behind the twisting somersault, which include a series of classical papers \cite{yeadon93a, yeadon93b, yeadon93c, yeadon93d}. In this paper we focus on optimising somersaults without twist, which we will refer to as the planar somersault. There are several studies focusing on the planar somersault, e.g.\! \cite{Blajer2003, King2004471, MurthyK93, schuler2011optimal}, but here we take a different approach that utilises the geometric phase in order to maximise somersault rotation.
The geometric phase is also important in the twisting somersault, see \cite{TwistSom}, and has been used to generate a new dive in \cite{513XD}. 
The main focus for the twisting somersault is the generation of twist, while in the planar case the particulars of the shape change can instead be used to generate additional somersault.
Our initial formulas are a special case of those derived in \cite{TwistSom}, but since the resulting differential equation for overall rotation is only 
a single first order equation, a much more thorough analysis is possible.

The splitting into the dynamic and geometric phases is a well known modern concept in geometric mechanics, 
see e.g.\! \cite{MarsdenRatiu13,Holm08,Holm09}. To our knowledge this is the first application of 
geometric phase to diving. When modelling the rotation of a non-rigid 
(and hence a shape changing) body the contributions to the overall rotation 
can be split into two terms.
The more familiar term is the dynamic phase, which is proportional to the angular momentum of the body and thus expresses the obvious fact that the body rotates faster when it has larger angular momentum.
The less familiar term is {\em not} proportional to the angular momentum, and is only present 
when a shape change occurs. This term is called the geometric phase, and it gives a contribution 
to the overall rotation of the body even when the angular momentum vanishes. This is the 
reason a cat can change orientation by changing its body shape even when it has no angular momentum.
In diving angular momentum is non-zero, but the geometric phase is still present, and both 
together determine the overall rotation.

The structure of the paper is as follows. In section \ref{sec:model} we take the mathematical model of an athlete introduced in \cite{WTongthesis}, and simplify it to analyse planar somersaults. In section \ref{sec:eqofmotion} we then present the generalised equations of motion for coupled rigid bodies in space, and perform planar reduction to reduce the 3-dimensional vector equations into the 2-dimensional scalar variants. Next in section \ref{sec:realworld} we analyse a real world dive and demonstrate how the geometric phase can be used to improve overall rotation obtained by the athlete. Finally, in section \ref{sec:theoreticalplanar} we propose a new theoretical planar somersault using realistic assumptions to find optimal shape change trajectories that maximise overall rotation. In these instances the dynamic and geometric phases are accessed for different values of angular momentum, and the role of the geometric phase in optimising overall rotation is demonstrated.

A preliminary version of this study was presented at the 1st Symposium for Researchers in Diving at Leipzig, Germany. The conference proceedings can be found in \cite{divingsym}. However, here the model presented in section 2 has been tweaked, the analysis of the real world dive in section 4 has been extended to show how the geometric phase can be utilised to increase the amount of somersault produced, and we present a new optimisation procedure in section 5 that maximises the theoretical planar somersault using the geometric phase.

\section{Model}\label{sec:model}
We begin with the 10-body model proposed in \cite{WTongthesis}, which is a slight modification of Frohlich's \cite{Frohlich} 12-body model that uses simple geometric solids connected at joints to represent the athlete. The components of the 10-body model consist of the torso, head, $2\times$ upper arm, $2\times$ forearm with hand attached, $2\times$ thigh and $2\times$ lower leg with foot attached. Although more sophisticated models exist (like those proposed by Jensen \cite{Jensen76} and Hatze \cite{Hatze}) that use the elliptical zone method to estimate segment parameters, this sophistication only affects the tensor of inertia $I_i$, centre of mass $\bo{C}_i$ and joint location $\bo{E}_i^j$ for each body $B_i$ connected to body $B_j$. As the dynamics are driven only by $I_i$, $\bo{C}_i$ and $\bo{E}_i^j$, we can use the 10-body model since it provides similar estimates for these quantities. 

\begin{figure}[b]
\centering
\subfloat[Body segments of the model.]{\includegraphics[width=7.25cm]{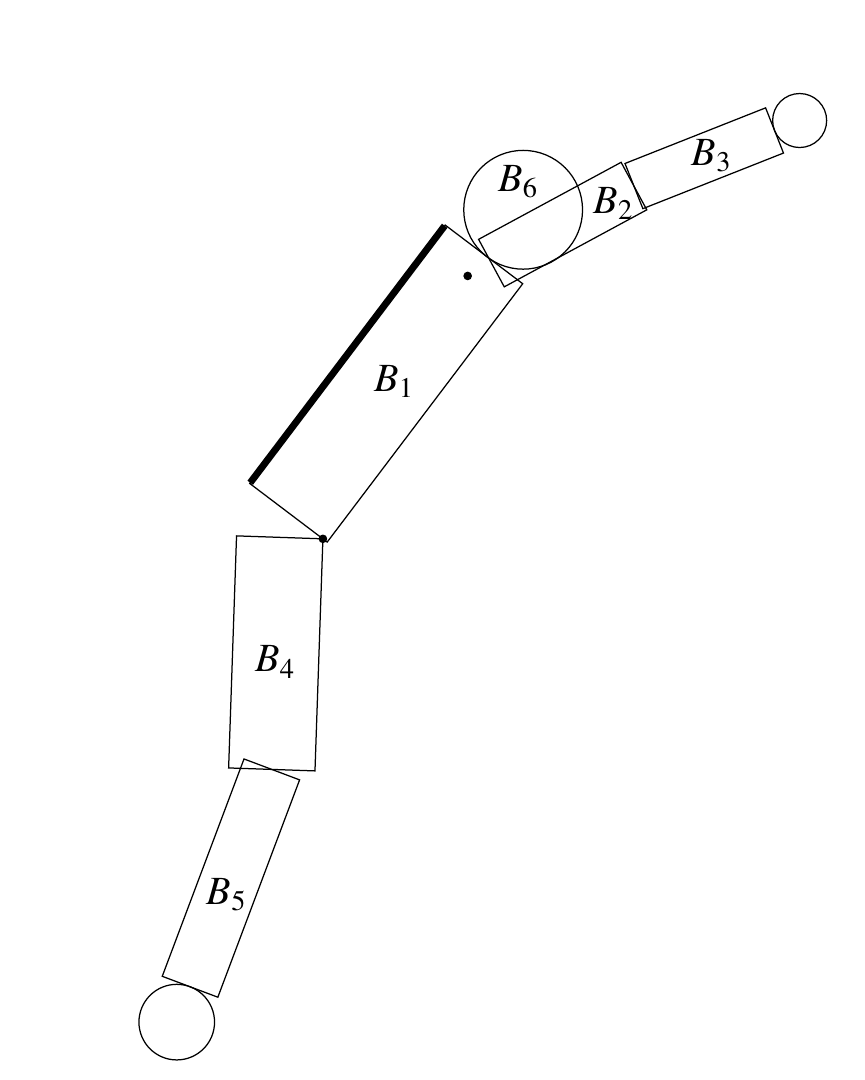}}
\hspace*{3mm}\subfloat[Joint vectors of the model.]{\includegraphics[width=7.25cm]{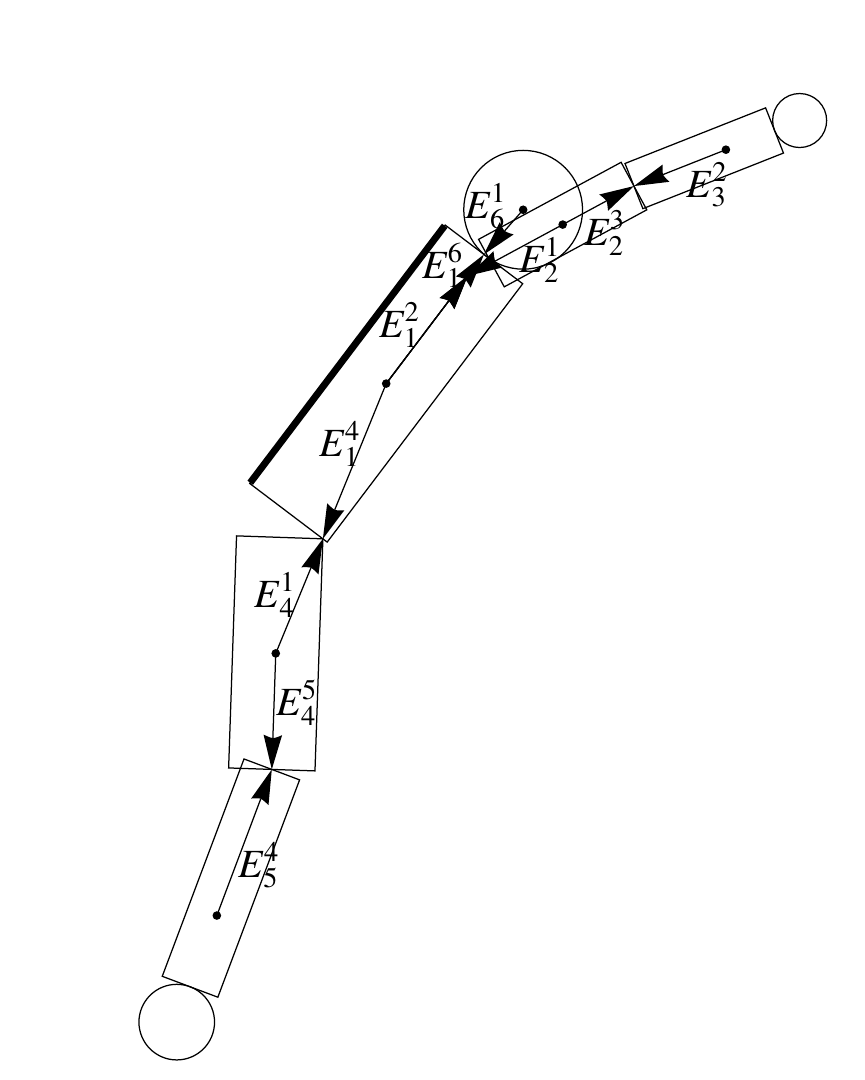}}
\caption{The schematic diagrams illustrate the collection of body segments $B_i$ and joint vectors $\bo{E}_i^j$. Each joint vector $\bo{E}_i^j$ shows the connection between $B_i$ and $B_j$, and is a constant vector when measured in the $B_i$-frame.}
\label{fig:schematic}
\end{figure}

In the case of the planar somersault we enforce the shape change to be strictly about the somersault axis and require that both left and right limbs move together, so that the normalised angular velocity vector is constant. By combining the corresponding left and right limb segments we obtain the 6-body planar model shown in Figure \ref{fig:schematic}, whose parameters are listed in Table \ref{tab:model}.

\begin{table}[t]\centering\begin{tabular}{|l|l|l|l|}
\hline
body $B_i$ & mass $m_i$ & moi $I_i$ & joint position $\bo{E}_i^j$ \\\hline
$B_1=$ torso						 & $m_1=32.400$ & $I_1=1.059$ & $\bo{E}_1^2=(0,0.25)^t$\\
												 &						  &							& $\bo{E}_1^4=(0.08,-0.3)^t$\\
												 &						  &							& $\bo{E}_1^6=(0,0.3)^t$\\
$B_2=$ upper arms				 & $m_2=4.712$  & $I_2=0.038$ & $\bo{E}_2^1=(0,0.2)^t$\\
												 &						  &							& $\bo{E}_2^3=(0,-0.15)^t$\\
$B_3=$ forearms \& hands & $m_3=4.608$  & $I_3=0.055$ & $\bo{E}_3^2=(0,0.183)^t$\\
$B_4=$ thighs						 & $m_4=17.300$ & $I_4=0.294$ & $\bo{E}_4^1=(0.08,0.215)^t$\\
												 &						  &			 			  & $\bo{E}_4^5=(0,-0.215)^t$\\
$B_5=$ lower legs \& feet& $m_5=11.044$ & $I_5=0.310$ & $\bo{E}_5^4=(0.08,0.289)^t$\\
$B_6=$ head							 & $m_6=5.575$  & $I_6=0.027$ & $\bo{E}_6^1=(0,-0.11)^t$\\
\hline
\end{tabular}
\caption{The parameters of the 6-body planar model obtained by reducing the 10-body model proposed in \cite{WTongthesis}. We abbreviate moment of inertia as moi in the table above.}
\label{tab:model}
\end{table}

\section{Equations of motion}\label{sec:eqofmotion}
The equations of motion for a rigid body in 3-dimensional space is 
\begin{equation}
\bo{\dot{L}}=\bo{L}\times \bo{\Omega},\label{eq:eom}
\end{equation}
where $\bo{L}$ is the angular momentum and $\bo{\Omega}$ is the angular velocity. Now if the rigid body is replaced by a system of coupled rigid bodies, then \eqref{eq:eom} holds if
\begin{equation}
\bo{\Omega} = I^{-1}(\bo{L}-\bo{A}).\label{eq:omega}
\end{equation}
Here, $I$ is the overall tensor of inertia and $\bo{A}$ is the total momentum shift generated by the shape change. Thus in the absence of shape change $\bo{A}=\bo{0}$, which gives the classical result $\bo{\Omega} = I^{-1}\bo{L}$. The proof of \eqref{eq:omega} is provided in Theorem 1 in \cite{TwistSom}, along with $I$ and $\bo{A}$, which are
\begin{align}
I&=\sum_{i=1}^6\Big(R_{\alpha_i} I_i R_{\alpha_i}^t+m_i\left[|\bo{C}_i|^2\mathbb{1}-\bo{C}_i\bo{C}_i^t\right]\Big)\label{eq:I}\\
\bo{A}&=\sum_{i=1}^6\Big(m_i \bo{C}_i\times \bo{\dot{C}}_i +R_{\alpha_i} I_i\bo{\Omega}_{\alpha_i}\Big).\label{eq:A}
\end{align}
In \eqref{eq:I} and \eqref{eq:A} for each body $B_i$ we have: the mass $m_i$, the tensor of inertia $I_i$, the centre of mass $\bo{C}_i$, the relative orientation $R_{\alpha_i}$ to the reference body (chosen to be $B_1$), and the relative angular velocity $\bo{\Omega}_{\alpha_i}$, such that the angular velocity tensor is $\hat{\Omega}_{\alpha_i} = R_{\alpha_i}^t \dot{R}_{\alpha_i}$. 
It is clear that when there is no shape change $\dot{\bo{C}}_i$ and $\bo{\Omega}_{\alpha_i}$ 
both vanish, and hence $\bo{A}$ vanishes. 
Manipulating \eqref{eq:A} by using the definition of $\bo{C}_i$ found in \cite{WTongthesis} and the vector triple product formula, we can factorise out the relative angular velocity $\bo{\Omega}_{\alpha_i}$ to obtain
\begin{equation}
\bo{A}=\sum_{i=1}^6\Big(R_{\alpha_i}\Big[\sum_{j=1}^6 m_j [R_{\alpha_i}^t \bo{C}_j \cdot \bo{\tilde{D}}_i^j \mathbb{1} - \bo{\tilde{D}}_i^j \bo{C}_j^t R_{\alpha_i}] + I_i \Big]\bo{\Omega}_{\alpha_i}\Big),\label{eq:A2}
\end{equation}
where $\bo{\tilde{D}}_i^j$ is a linear combination of $\bo{E}_i^j$'s defined in Appendix \ref{app:D}.

For planar somersaults the angular momentum and angular velocity vector are only non-zero about the somersault axis, so we write $\bo{L}=(0,L,0)^t$ and $\bo{\Omega}=(0,\dot{\theta},0)^t$ to be consistent with \cite{WTongthesis}. Substituting $\bo{L}$ and $\bo{\Omega}$ in \eqref{eq:eom} shows that $\bo{\dot{L}}=\bo{0}$, meaning the angular momentum is constant. As the athlete is symmetric about the median plane, the centre of mass for each $B_i$ takes the form $\bo{C}_i = (C_{i_x},0,C_{i_z})^t$ and the relative angular velocity $\bo{\Omega}_{\alpha_i}=(0,\dot{\alpha}_i,0)^t$. Substituting these results in \eqref{eq:I} and \eqref{eq:A2}, we can then simplify to obtain the 2-dimensional analogue of $I$ and $\bo{A}$ giving
\begin{align}
I^{(2D)} &= \sum_{i=1}^6 I_{i_y}+m_i(C_{i_x}^2+C_{i_z}^2)\\
A^{(2D)} &= \sum_{i=1}^6\Big( \sum_{j=1}^6 m_j\big[\bo{C}_j\cdot \bo{\tilde{D}}_i^j \cos{\alpha_i} + P\bo{C}_j\cdot \bo{\tilde{D}}_i^j \sin{\alpha_i}\big]+I_{i_y}\Big)\dot{\alpha}_i,
\end{align}
where $I_{i_y}$ is the (2,2) entry of $I_i$ and $P = \text{antidiag}(-1,1,1)$ is an anti-diagonal matrix used to permute the components of $\bo{C}_j$. 
The 2-dimensional analogue of \eqref{eq:omega} with the arguments $\bo{\alpha}=(\alpha_2,\dots,\alpha_6)^t$ (note $\alpha_1=0$ because $B_1$ is the reference segment) explicitly written and $(2D)$ superscripts suppressed is then
\begin{equation} 
\dot{\theta} = I^{-1}(\bo{\alpha})L+\bo{F}(\bo{\alpha})\cdot\bo{\dot{\alpha}},\label{eq:theta}
\end{equation}
where we write $\bo{F}(\bo{\alpha})\cdot\bo{\dot{\alpha}} = -I^{-1}(\bo{\alpha})A(\bo{\alpha},\bo{\dot{\alpha}})$ to match the differential equation found in \cite{pentagon}. In that paper $L=0$, and the study focused on maximising the geometric phase. However, here $L\neq 0$ and the differential equation \eqref{eq:theta} is composed of two parts: the dynamic phase $I^{-1}(\bo{\alpha})L$, which is proportional to $L$, and the geometric phase $\bo{F}(\bo{\alpha})\cdot\bo{\dot{\alpha}}$, which is independent of $L$. Solving \eqref{eq:theta} with the initial condition $\theta(0)=\theta_0$ gives the orientation of the athlete as a function of time. 

The sequence of shape changes an athlete goes through while performing a dive can be represented by a curve on shape space, which closes into a loop provided the athlete's take-off and final shape is the same. The dynamic and geometric phases are both dependent on the path of the loop, however the dynamic phase also depends on the velocity with which the loop is traversed, while the geometric phase is independent of the velocity. Traversing the same loop with different velocities therefore contributes different amounts to the dynamic phase, while the contribution to the geometric phase is unchanged. For planar somersaults, we generally expect the dominating term to be the dynamic phase as it is proportional to $L$ (which is large), and the geometric phase to play a lesser role.
 
The main idea behind our optimisation is that we assume $L$ is already as large as possible
and cannot be increased further. Also, we assume that the athlete is holding tuck or pike as tight as possible in the middle of the dive, so again no further improvement is possible. Both increasing $L$ and decreasing $I$ change the dynamic phase, which increases overall rotation and thus the
number of somersaults. We will show that the contribution from the second term in \eqref{eq:theta},
the geometric phase, can be used in principle to increase the overall rotation.

\section{A real world dive}\label{sec:realworld}
\begin{figure}[t]
\centering
\includegraphics[width=14cm]{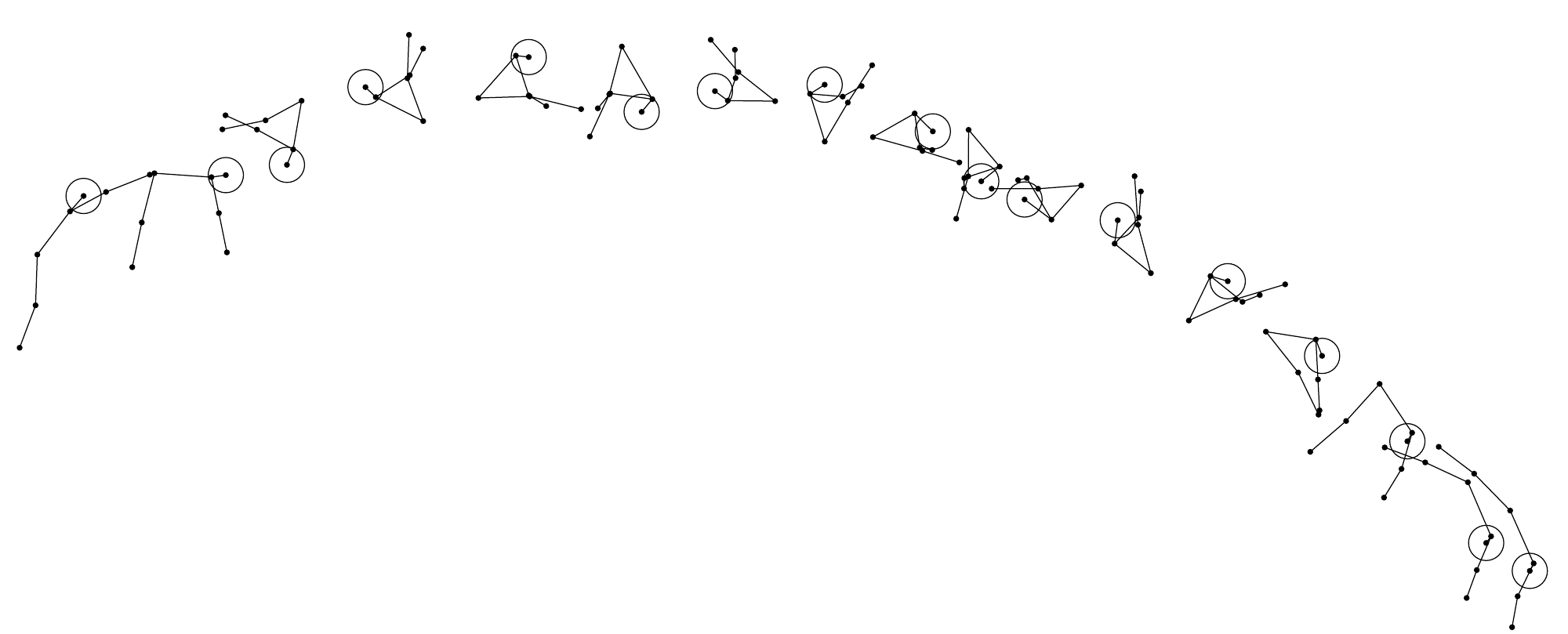}
\caption{Illustration of the digitised dive for frames 0, 15, 30, 45, 60, 75, 90, 100, 110, 120, 130, 140, 150, 160, 170, 180 and 186, from left to right. To avoid clutter, each illustrated frame has been shifted right by a small constant amount to provide better visualisation of the dive sequence.}\label{fig:digitise}
\end{figure}

Footage of a professional male athlete performing a 107B dive (forward 3.5 somersaults in pike) off the 3m springboard was captured at the New South Wales Institute of Sport (NSWIS) using a 120 FPS camera. SkillSpector \cite{skillspector} was used to manually digitise the footage, which commenced from the moment of take-off and ended once the athlete's hand first made contact with the water upon entry. The total airborne time spanned 1.55 seconds, creating a total of 187 frames, and the digitisation of the dive is shown in Figure \ref{fig:digitise}. For convenience we shall refer to the initial frame as the zeroth frame, and write $\bo{\alpha}[j]=\{\alpha_2[j],\dots\}$ to denote the collection of shape angles of the $j$th frame, where $0 \leq j \leq 186$.

\begin{figure}[p]
\centering
\subfloat[Orientation of the athlete.]{\includegraphics[width=7.25cm]{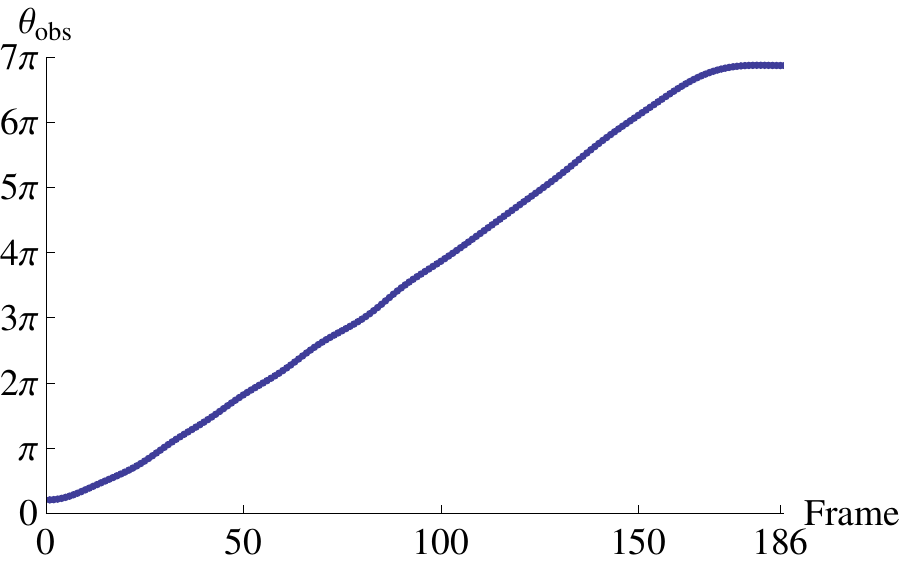}}
\hspace*{2mm}\subfloat[Relative angle of the upper arms.]{\includegraphics[width=7.25cm]{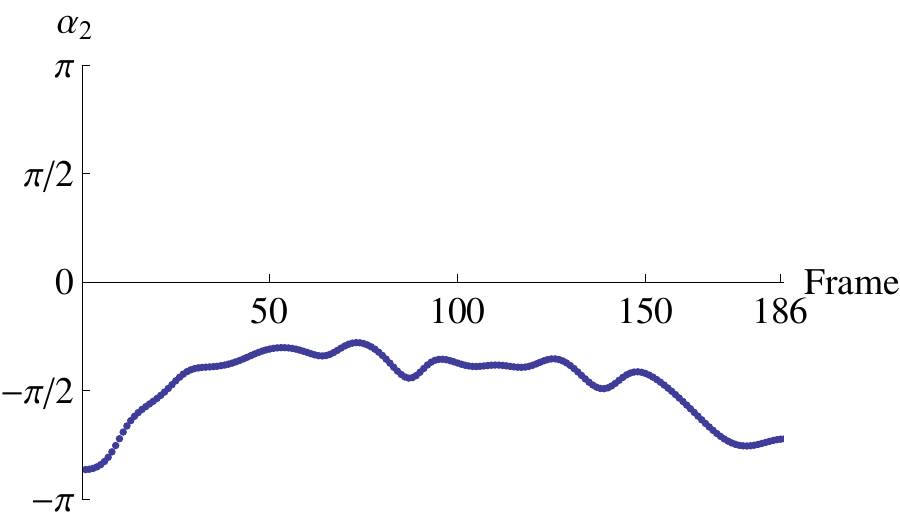}}\\
\subfloat[Relative angle of the forearms and hands.]{\includegraphics[width=7.25cm]{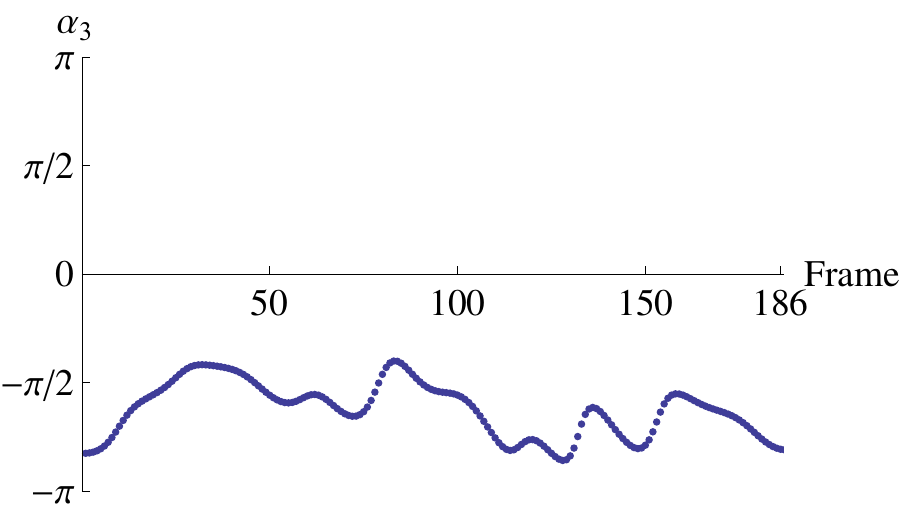}}
\hspace*{2mm}\subfloat[Relative angle of the thighs.]{\includegraphics[width=7.25cm]{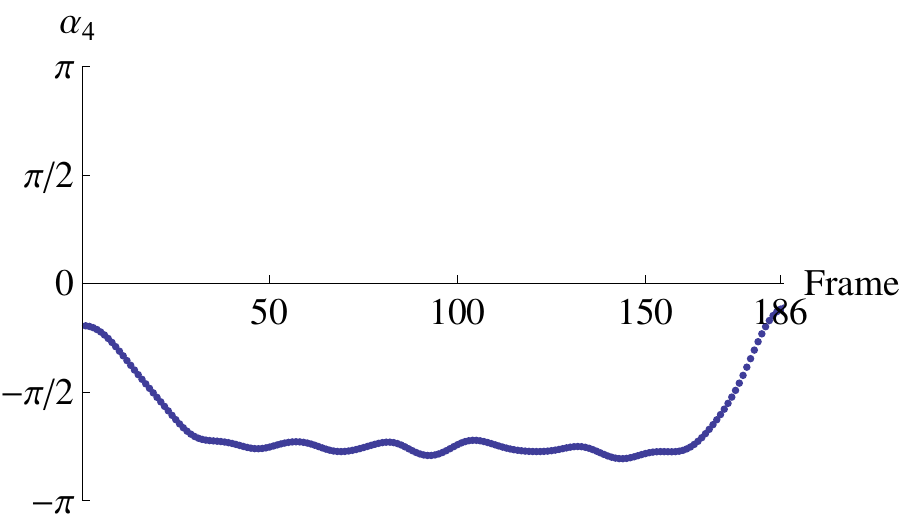}}\\
\subfloat[Relative angle of the lower legs and feet.]{\includegraphics[width=7.25cm]{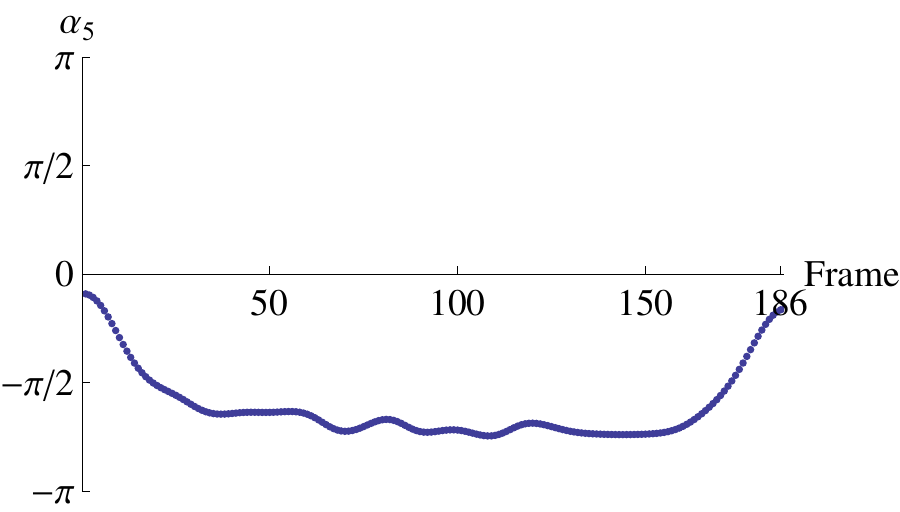}}
\hspace*{2mm}\subfloat[Relative angle of the head.]{\includegraphics[width=7.25cm]{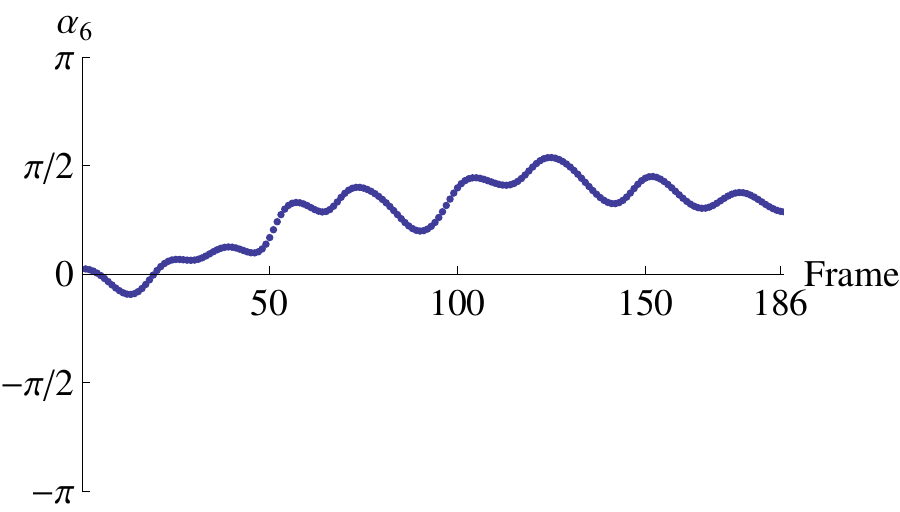}}
\caption{The orientation of the athlete is specified by $\theta_\mathit{obs}$ and the collection of $\alpha_i$ specifies the shape.}
\label{fig:angles}
\end{figure}

In each frame we locate the joint positions of the ankle, knee, hip, shoulder, elbow, wrist, and ear (which serves as a decent approximation for the centre of mass of the head). To reduce digitisation errors a discrete Fourier cosine transform is applied to the data, 
so that by keeping the first fifteen Fourier coefficients the data is smoothed when inverting the transformation. The spatial orientation $\theta_i$ is the angle between an orientation vector constructed from appropriate joint positions (e.g.\! hip and shoulder positions for the torso) and the reference vector given by the anatomical neutral position vector when standing upright. From $\theta_i$ the relative orientation $\alpha_i$ can be obtained by using  
$\alpha_i = \theta_i - \theta_\mathit{obs}$ for $i\in\{2,\dots,6\}$, where $\theta_\mathit{obs}=\theta_1$ is the observed spatial orientation of the torso. The spatial orientation $\theta_\mathit{obs}$ and relative orientations $\alpha_2,\dots,\alpha_6$ for the digitised dive are shown in Figure \ref{fig:angles}.

\begin{figure}[t]
\centering
\includegraphics[width=9cm]{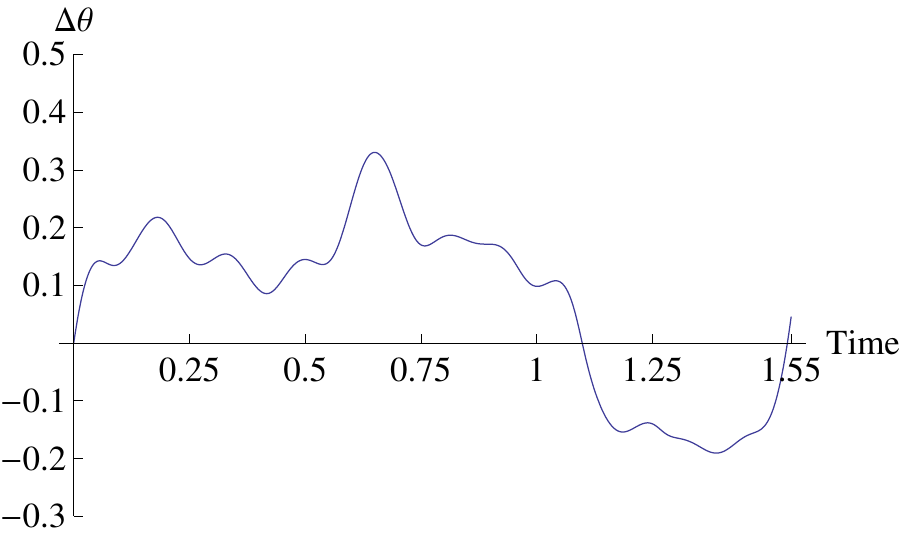}
\caption{The difference between $\theta$ computed with \eqref{eq:theta} and the observed $\theta_\mathit{obs}$ from the data.}\label{fig:deltatheta}
\end{figure}

To validate the 6-body planar model we compute $\theta$ using \eqref{eq:theta} and compare the theoretical result to the observed $\theta_\mathit{obs}$. Small variations between the two curves are expected as the segment parameters are taken from Table \ref{tab:model} and not from the particular athlete used in the data. Comparison is made by using the initial orientation 
\begin{equation}
\theta_0=\theta_\mathit{obs}[0]=0.6478\label{eq:theta0}
\end{equation}
and collection of shape angles $\bo{\alpha}[\mathit{fr}]$, where square brackets denote frame $\mathit{fr}$ and round brackets indicate continuous time $t$. A cubic interpolation is used to obtain $\bo{\alpha}(t)$, which is then substituted in \eqref{eq:theta} to obtain $\theta$. The angular momentum constant $L$ is found by a least square fit that reveals $L=122.756$, and the difference $\Delta\theta=\theta-\theta_\mathit{obs}$ is shown in Figure \ref{fig:deltatheta}. From the moment of take-off until the diver hits the water we see that the discrepancy remains small, with the maximum difference being $0.331$ radians at time $t=0.650$. When the dive is completed the difference is only $0.045$ radians, which gives us confidence in using the planar model.
\begin{figure}[b]
\centering
\includegraphics[width=5cm]{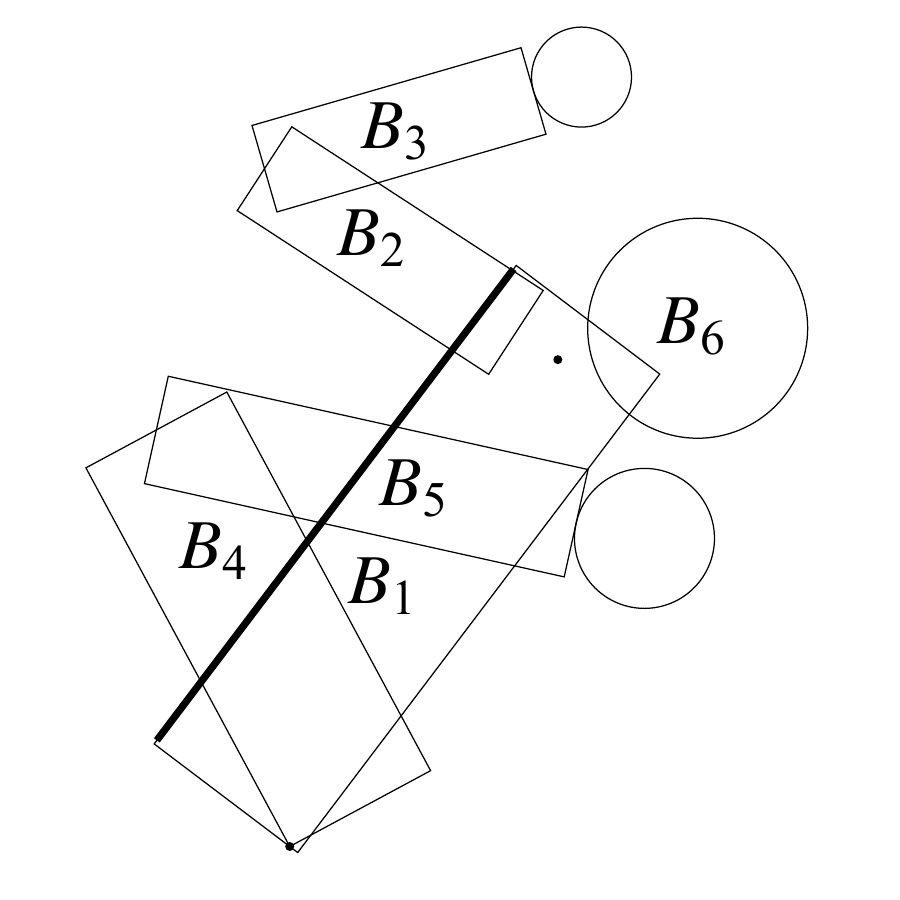}
\caption{An impossible shape configuration of the athlete with $\theta = 0.648$ and $\bo{\alpha}=\{1.5,-2.5,2,-2,1\}$.}\label{fig:weird}
\end{figure}

Looking at the graphs in Figure \ref{fig:angles} it appears the upper arms and forearms move together relative to the torso, and similarly the thighs and lower legs. Therefore it is reasonable to assume that the elbows and knees remain straight throughout the dive, in particular the knees during pike somersaults. The head can be included 
in the torso to remove another degree of freedom. Using these assumptions thus further reduces the segment count to three so that the shape space is some subset of a 2-dimensional torus. We say subset as there are shapes deemed impossible or unrealistic for diving, e.g.\! the shape shown in Figure \ref{fig:weird}. This motivates us to constrain the shape space to $\{\alpha_2,\alpha_4\}\in[-\pi,0]^2$ for the set of all possible shapes obtainable by the athlete. An important point to note is that the general theory is not dependent on the segment count reduction, and that these assumptions merely make it easier to understand the main principles behind the theory. Using $L=120$ for comparison purposes, we found the overall difference in rotation obtained to be less than $6.53^\circ$, which is small considering the athlete completes 3.5 somersaults for the dive.

\begin{figure}[t]
\centering
\hspace*{-10mm}\subfloat[The full trajectory.]{\includegraphics[width=7.8cm]{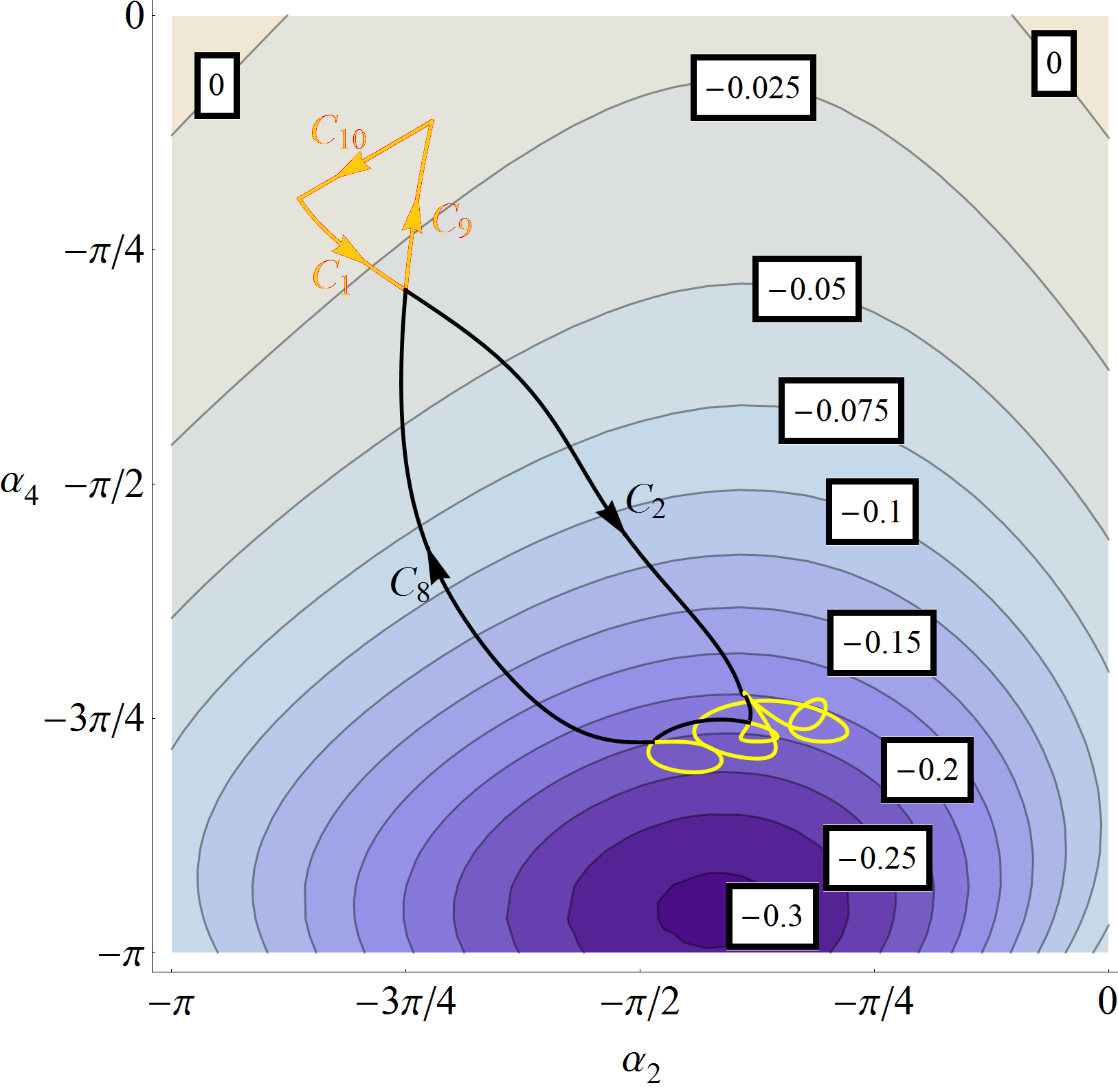}}
\hspace{2mm}\subfloat[Close-up of the left pane.]{\includegraphics[width=8.1cm]{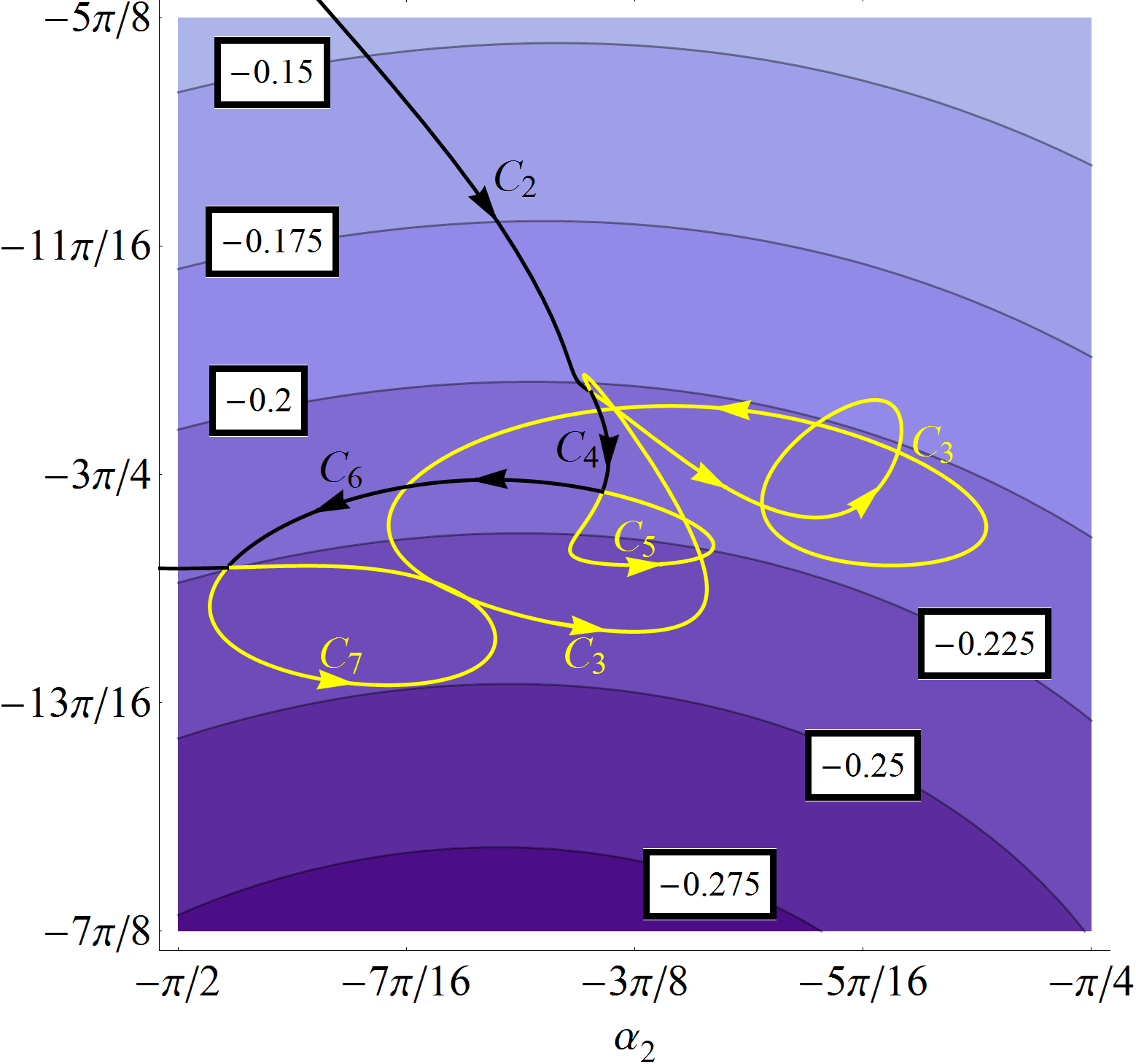}}
\caption{The interpolated shape change trajectory plotted with constant $B$ contours. The 10 pieces $C_1,\dots, C_{10}$ of $C$ are also identified here.}\label{fig:shapespace}
\end{figure}

\begin{figure}[b]
\centering
\hspace*{-0mm}\subfloat[The shape trajectories of the original dive where the blue regions denote sub-loops that require their directions reversed to maximise the geometric phase.]{\includegraphics[width=16cm]{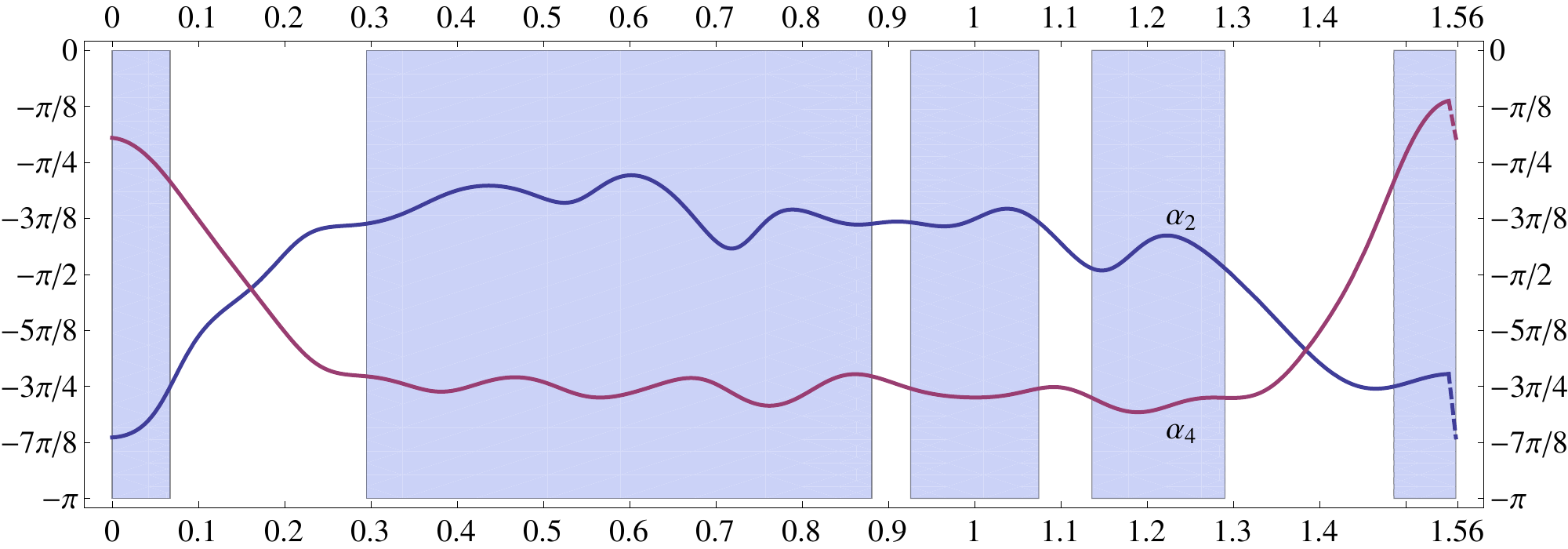}}\\
\hspace{0mm}\subfloat[The shape trajectories after orientating all sub-loops to provide a positive contribution to the geometric phase. The vertical dashed lines indicate where the pieces are joined.]{\includegraphics[width=16cm]{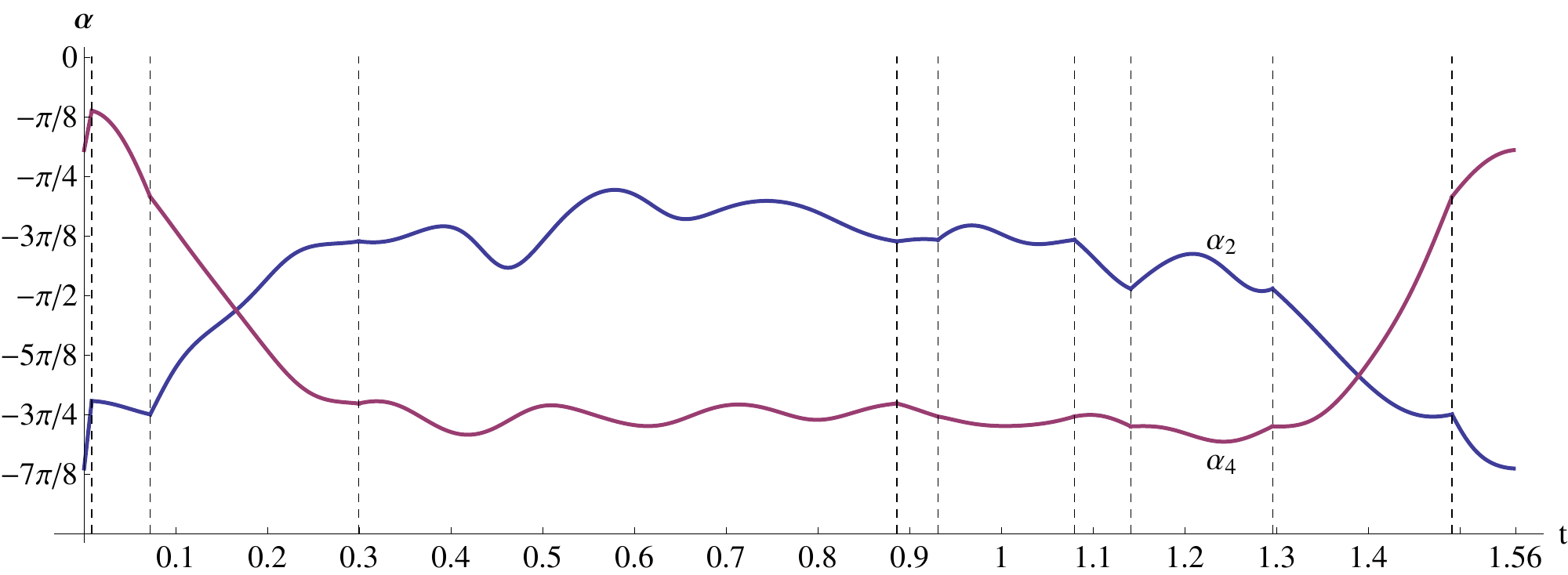}}
\caption{The shape trajectories before and after orientating the sub-loops.}\label{fig:loops}
\end{figure}

The athlete's 2-dimensional shape change trajectory is shown in Figure \ref{fig:shapespace}, where the loop $C$ is constructed by letting $\bo{\alpha}(t)$ run from zero to the airborne time $T_\mathit{air}$. As take-off and entry shapes are not identical, the overall rotation obtained is gauge dependent, see \cite{LittlejohnReinsch97} for details. To eliminate this ambiguity we add one additional frame where $\bo{\alpha}[187] = \bo{\alpha}[0]$, such that $\bo{\alpha}(t)$ closes and the overall rotation is well-defined. As we now have a total of 188 frames captured at 120 FPS, the additional frame adds $1/120$th of a second to the airborne time, thus making $T_\mathit{air} = 187/120$. To find the change in orientation of the dive we solve \eqref{eq:theta} with \eqref{eq:theta0} giving
\begin{equation}
\theta(T_\mathit{air})= \theta_\mathit{dyn} + \theta_\mathit{geo} + \theta_0,
\end{equation}
where the dynamic phase contribution is
\begin{equation}
\theta_\mathit{dyn} = L \int_0^{T_\mathit{air}} I^{-1}\big(\bo{\alpha}(t)\big)\,\mathrm{d}t= 0.1705L\label{eq:dyn}
\end{equation}
and the geometric phase contribution is
\begin{equation}
\theta_\mathit{geo} = \int_0^{T_\mathit{air}}\bo{F}\big(\bo{\alpha}(t)\big)\cdot\bo{\dot{\alpha}}(t)\,\mathrm{d}t= 0.0721.\label{eq:geo}
\end{equation}
As the angular momentum $L\approx 120$ is large, this confirms our hypothesis that the dynamic phase is the dominant term and the geometric phase only plays a minor role. For this particular dive the contribution is less than one degree of rotation, and hence is negligible.
However, we will show that this contribution can be increased to something tangible 
by changing the approach into and out of pike in the next section. 
Instead of directly integrating the second term in the differential equation \eqref{eq:theta}, the geometric phase can be obtained from certain areas in shape space.
If the loop in shape space is closed the integral can be rewritten in terms of a certain function over the enclosed area.
The reformulation is obtained by applying Green's theorem, which gives
\begin{equation}
\int_{C} F(\bo{\alpha})\cdot\dot{\bo{\alpha}}\,dt = \iint_A B(\bo{\alpha})\,d\tilde{A},\label{eq:green}
\end{equation}
where $A$ is the region enclosed by $C$ and $B(\bo{\alpha})$ is interpreted as the magnetic field with constant contours shown in Figure \ref{fig:shapespace}. 
The advantage of the area formulation is that it is easy to visualise the function $B(\alpha)$ in shape
space, and wherever this function is large the potential contribution to the geometric phase is large as well.
By writing out $\bo{F}(\bo{\alpha})=F_1(\bo{\alpha})\bo{i}+F_2(\bo{\alpha})\bo{j}$ explicitly, we have
\begin{align}
F(\bo{\alpha})\cdot\dot{\bo{\alpha}} &= F_1(\bo{\alpha})\,\frac{d\alpha_2}{dt}+F_2(\bo{\alpha})\,\frac{d\alpha_4}{dt} &\text{ and }&&
B(\bo{\alpha}) &= \frac{\partial F_2(\bo{\alpha})}{\partial \alpha_2}-\frac{\partial F_1(\bo{\alpha})}{\partial \alpha_4}\nonumber
\end{align}
appearing in \eqref{eq:green}. As $C$ is self-intersecting, the geometric phase can be computed by partitioning the loop into 10 pieces labelled $C_1,\dots, C_{10}$ (as shown in Figure \ref{fig:shapespace}) and then taking the appropriate pieces that make up the sub-loops. The total geometric phase is therefore the sum of the individual geometric phase contributions from each sub-loop. Specifically, the total geometric phase is composed of 
\begin{align}
(C_2,C_4,C_6,C_8) &= 0.0953 		& 		C_3 &= -0.0149		&		C_5 &= -0.0012\nonumber\\
(C_1,C_9,C_{10}) &= -0.0023			&			C_7 &= -0.0048,		&\nonumber
\end{align}
where summing the contributions from these five sub-loops yields $\theta_\mathit{geo}$ found in \eqref{eq:geo}.
We observe that most contributions towards the geometric phase are negative, meaning improvement can be made by reversing the direction of travel along the loop. As $B < 0$ throughout the dive, loops orientated clockwise will provide a positive contribution to the geometric phase, while loops oriented counterclockwise will provide a negative contribution.

We will now show that the geometric phase can be increased without changing the dynamic phase. 
Originally the pieces were ordered from $C_1$ to $C_{10}$, but after orientating the sub-loops the order becomes $-C_{10} \rightarrow -C_9 \rightarrow C_2\rightarrow -C_3\rightarrow C_4\rightarrow -C_5\rightarrow C_6\rightarrow -C_7\rightarrow C_8\rightarrow -C_1$, where a negative means we traverse along the piece in the opposite direction. The before and after shape trajectories are shown in Figure \ref{fig:loops}, where the improved geometric phase is
\begin{equation}
\tilde{\theta}_\mathit{geo} = 0.1186.
\end{equation}
While the effect from the geometric phase is small, it is still an improvement of $64\%$ compared to 
the original geometric phase \eqref{eq:geo}. 
Clearly these adjustments are not practical in an actual dive. However, we want to emphasise that we can increase the overall rotation achieved simply by reordering and reversing certain parts of the loop $C$. This additional rotation is obtained by maximising the geometric phase while leaving the dynamic phase unchanged.

\section{Optimising Planar Somersaults}\label{sec:theoreticalplanar}
\begin{figure}[t]
\centering
\includegraphics[width=14cm]{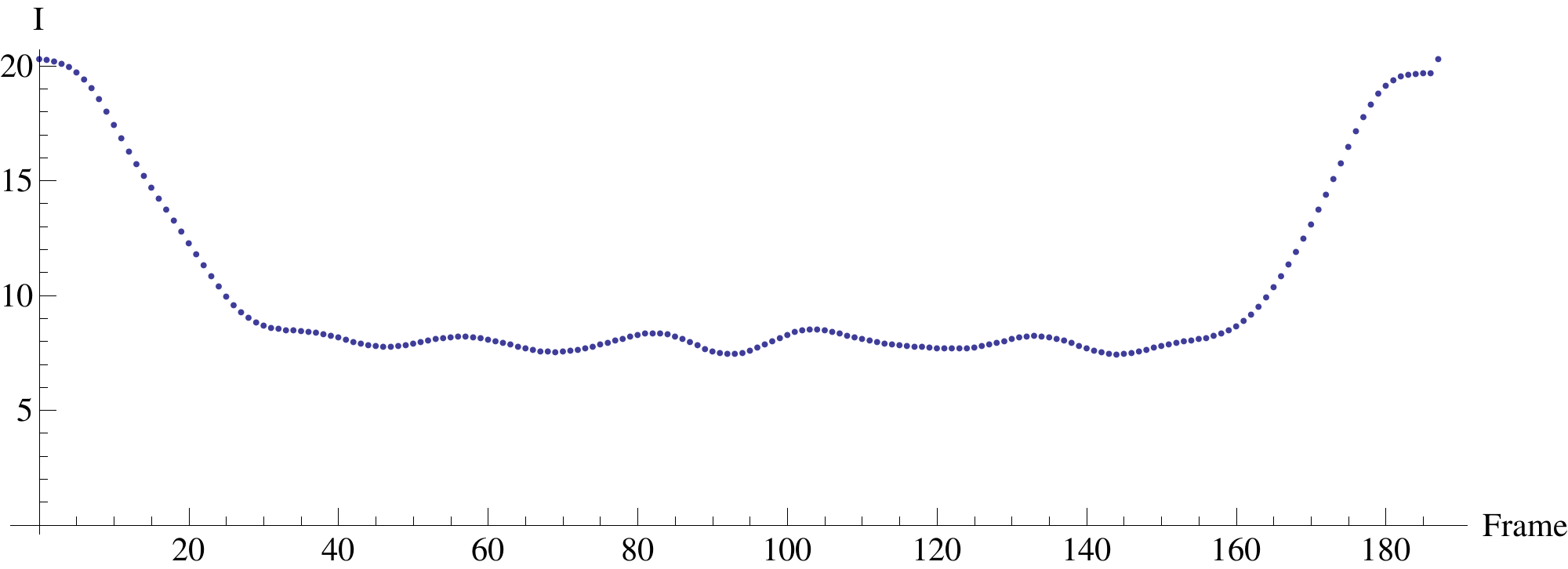}
\caption{The moment of inertia $I$ plotted frame by frame.}
\label{fig:moi}
\end{figure}
The following  observations can be made by analysing Figure \ref{fig:digitise} and Figure \ref{fig:moi}:
\begin{enumerate}
  \item[1.] The athlete takes off with a large moment of inertia.
  \item[2.] The athlete transitions quickly into pike position.
  \item[3.] The athlete maintains pike position during which the moment of inertia is small.
  \item[4.] The athlete completes the dive with (roughly) the same shape as during take-off.
\end{enumerate}
\begin{figure}[t]
\centering
\subfloat[Shape changing velocity of the arms.]{\includegraphics[width=15cm]{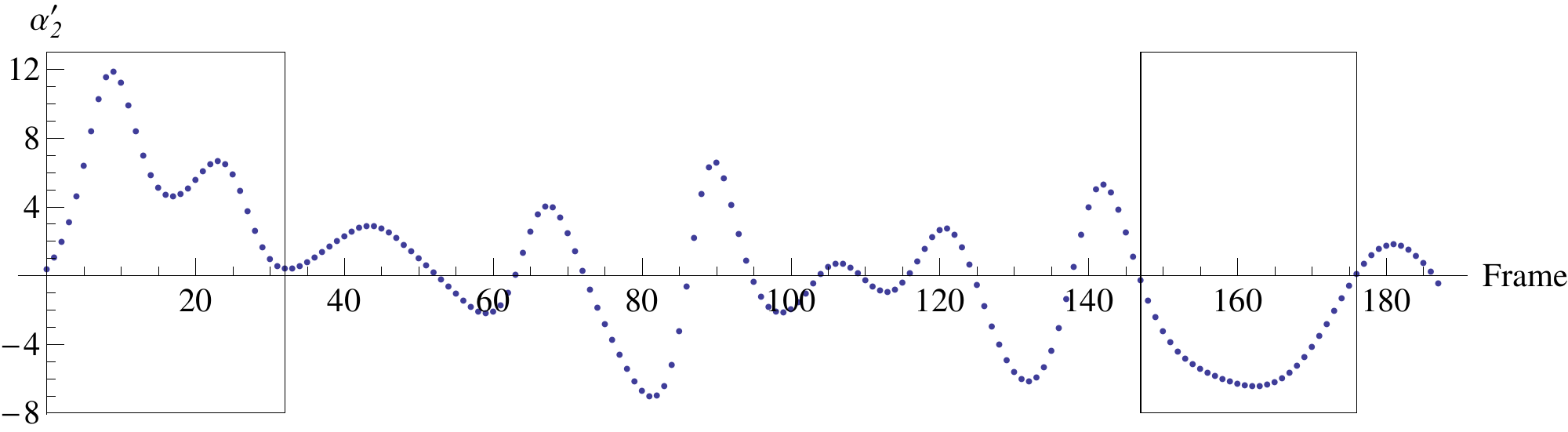}}\\
\subfloat[Shape changing velocity of the legs.]{\includegraphics[width=15cm]{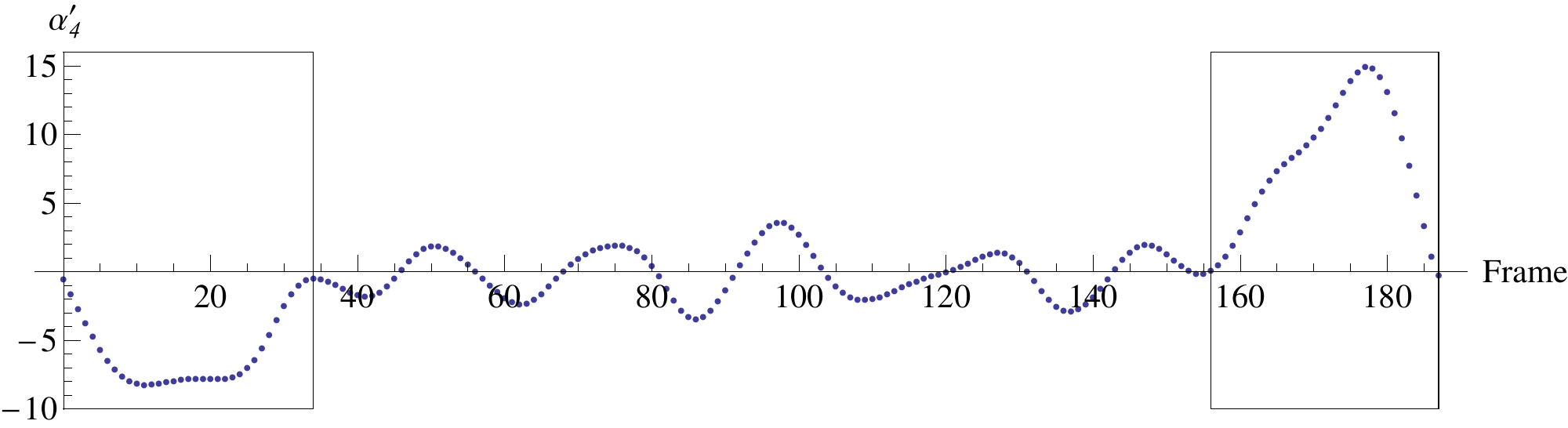}}
\caption{When entering pike position after take-off, the athlete's arms and legs reach their desired positions at frames 32 and 34, respectively. When leaving pike, the arms move first during frames 147 to 176, followed by the legs between frames 156 and 187.}
\label{fig:shapevelocity}
\end{figure}
Figure \ref{fig:shapevelocity} illustrates the relative velocities of the athlete's arms and legs, where the black boxes indicate transitions into and out of pike position that take at least a quarter second to complete. The interval between the two transition stages reveals small oscillations in the velocities, which are the result of the athlete making micro-adjustments to maintain pike position under heavy rotational forces.

\begin{figure}[t]
\centering
\subfloat[Shape $\bo{\alpha}_\mathit{max}$.]{\includegraphics[width=7.25cm]{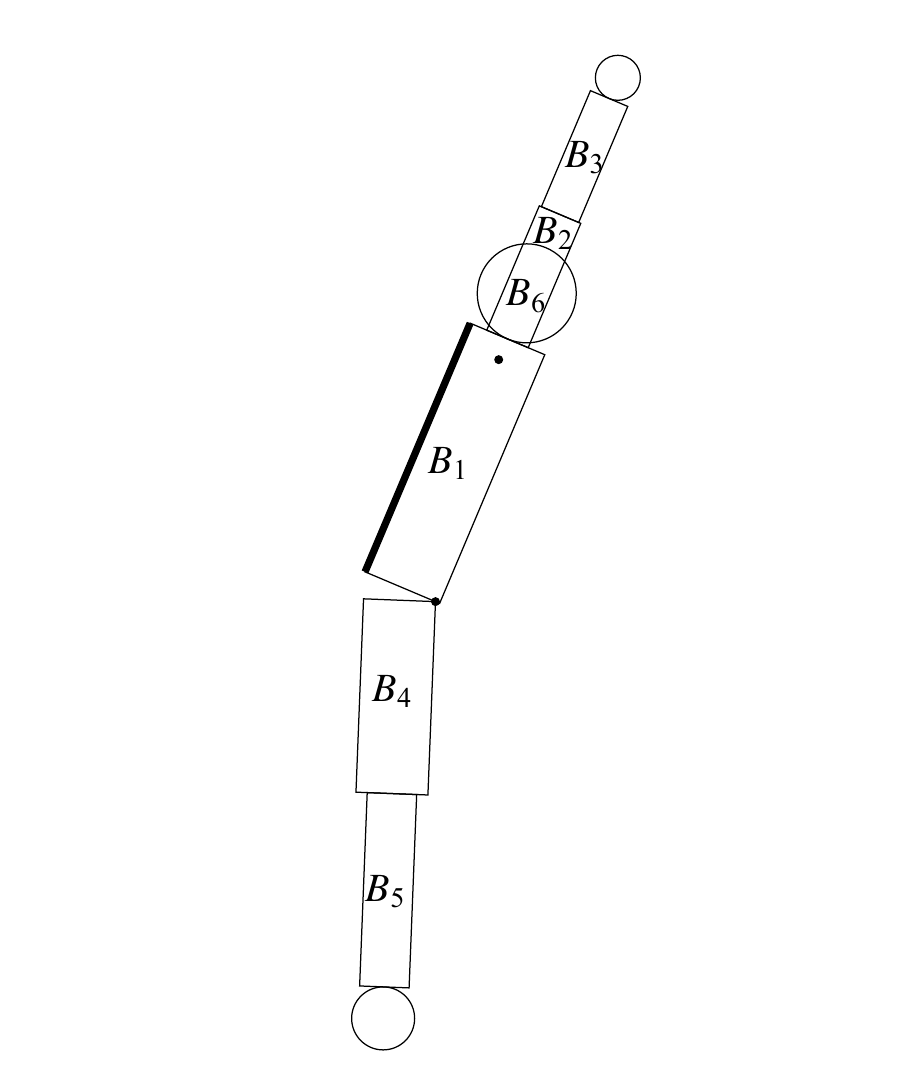}}
\subfloat[Shape $\bo{\alpha}_\mathit{min}$.]{\includegraphics[width=7.25cm]{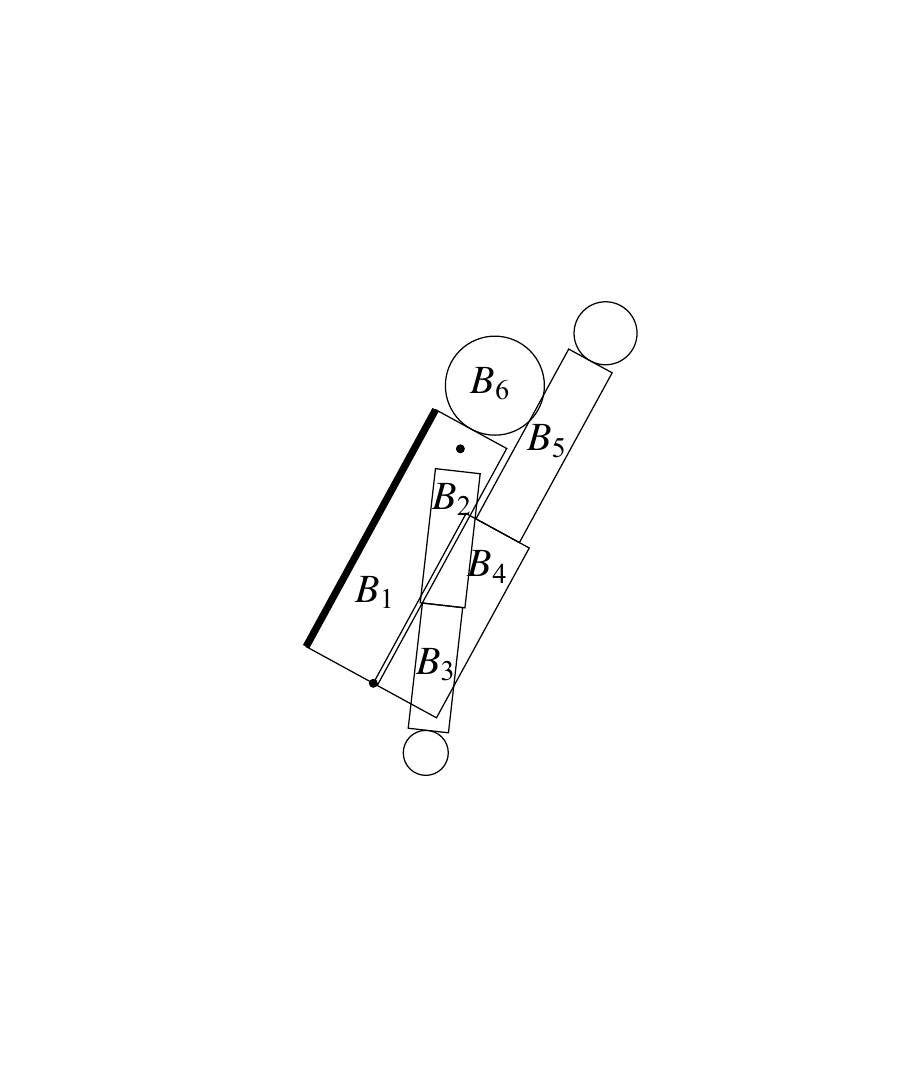}}
\caption{The shape configurations with extremum moments of inertia.}
\label{fig:Iextreme}
\end{figure}

Let $\bo{\alpha}_\mathit{max}\in[-\pi,0]^2$ correspond to the shape with maximum moment of inertia $I_\mathit{max}$, and let $\bo{\alpha}_\mathit{min}\in[-\pi,0]^2$ correspond to the shape with minimum moment of inertia $I_\mathit{min}$. We then find
\begin{align}
\bo{\alpha}_\mathit{max}&=(-\pi, -0.3608) & I_\mathit{max} &= 21.0647\nonumber\\
\bo{\alpha}_\mathit{min}&=(-0.3867,-\pi)  & I_\mathit{min} &= 5.2888\nonumber
\end{align}
and illustrate these shape configurations in Figure \ref{fig:Iextreme}. The figure reveals hip flexion in the shape corresponding to $\bo{\alpha}_\mathit{max}$, which is due to the positioning of the hip joint in the model. This resembles reality as athletes exhibit some hip flexion during take-off and entry into the water, as seen in Figure \ref{fig:digitise}.

\begin{figure}[t]
\centering
\includegraphics[width=12cm]{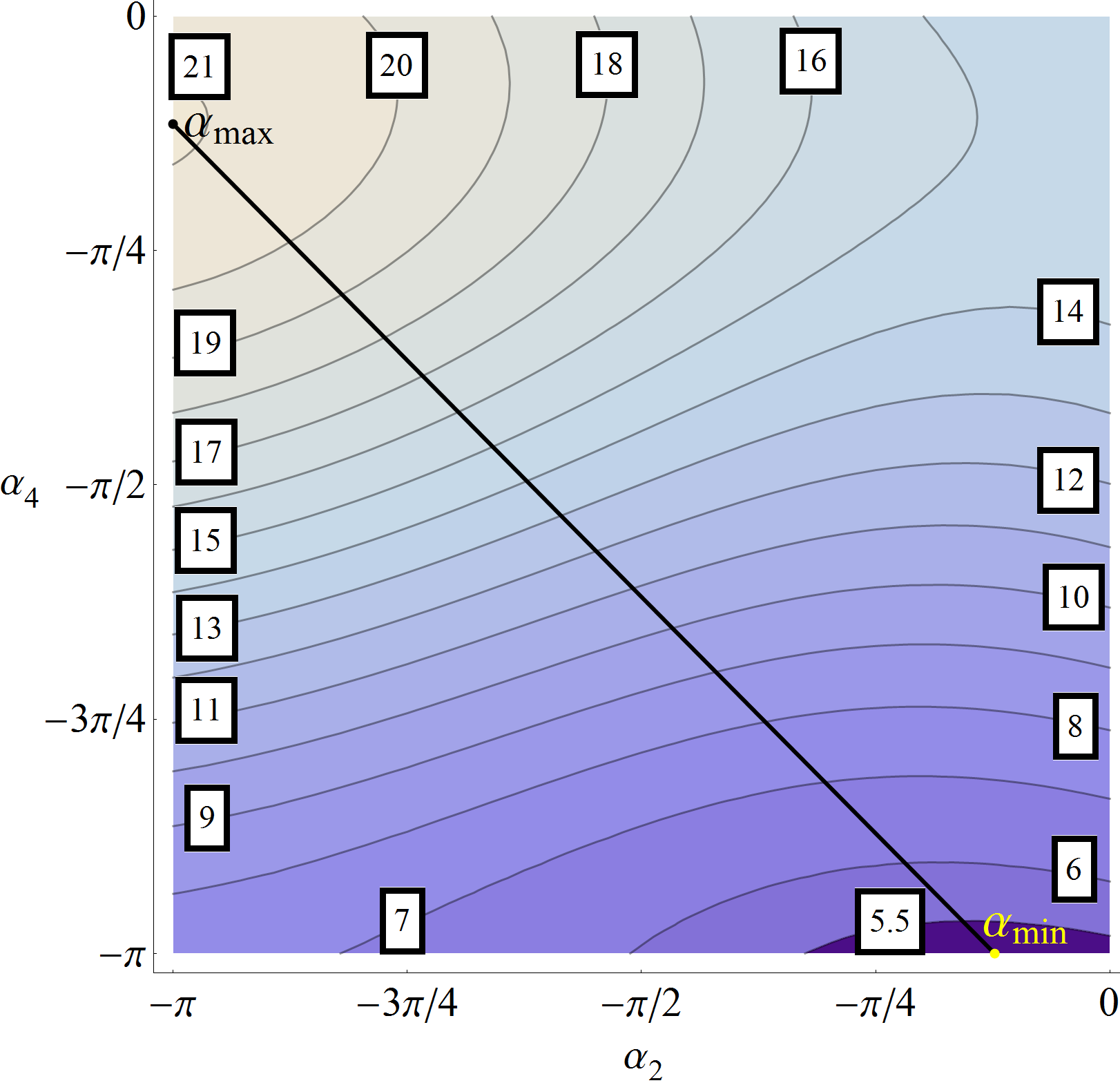}
\caption{The black curve illustrates the quickest way to move into and out of pike position so that the transition time is a quarter second. Constant $I(\bo{\alpha})$ contours have been plotted, which show that at $\bo{\alpha}_\mathit{max}$ and $\bo{\alpha}_\mathit{min}$ the moments of inertia are $I_\mathit{max}$ and $\bo{\alpha}_\mathit{min}$, respectively.}\label{fig:fastest}
\end{figure}

We now propose a theoretical dive by using the structure observed in the digitised dive as a guideline. In the idealised dive the athlete takes off with shape $\bo{\alpha}_\mathit{max}$ and immediately transitions into pike position specified by shape $\bo{\alpha}_\mathit{min}$. The athlete maintains pike without oscillations, then reverts back into the original shape $\bo{\alpha}_\mathit{max}$ to complete the dive. 
It appears obvious that this process will yield the maximal amount of somersault. However, as we will show, this is only true when the transition from $\bo{\alpha}_\mathit{max}$ to $\bo{\alpha}_\mathit{min}$ is instantaneous, which is of course unrealistic. When a maximum speed of shape change is imposed we will show that the amount of 
somersault can be increased slightly by using a different manoeuvre. The explanation behind this surprising observation is the geometric phase.
We cap $|\dot{\alpha}_2| = 11.0194$ (arms) and $|\dot{\alpha}_4| = 11.1230$ (legs)
when the limbs move at maximum speed, so that the transition from $\bo{\alpha}_\mathit{max}$ to $\bo{\alpha}_\mathit{min}$ (and vice versa) takes precisely a quarter second. Moving the limbs at maximum speed into and out of pike position maximises the time spent in pike, but there is no contribution to the amount of somersault from the geometric phase. We know the geometric phase is zero because there is no area enclosed by the loop $C$, as illustrated in Figure \ref{fig:fastest}.
The loop $C$ appears to be a line (and hence has no enclosed area) because the motions into pike and back into the layout position are happening in exactly the same way.
The essential idea is to break this symmetry by making the shape change into pike and the shape change back into the layout position different. 
\begin{figure}[t]
\centering
\includegraphics[width=12cm]{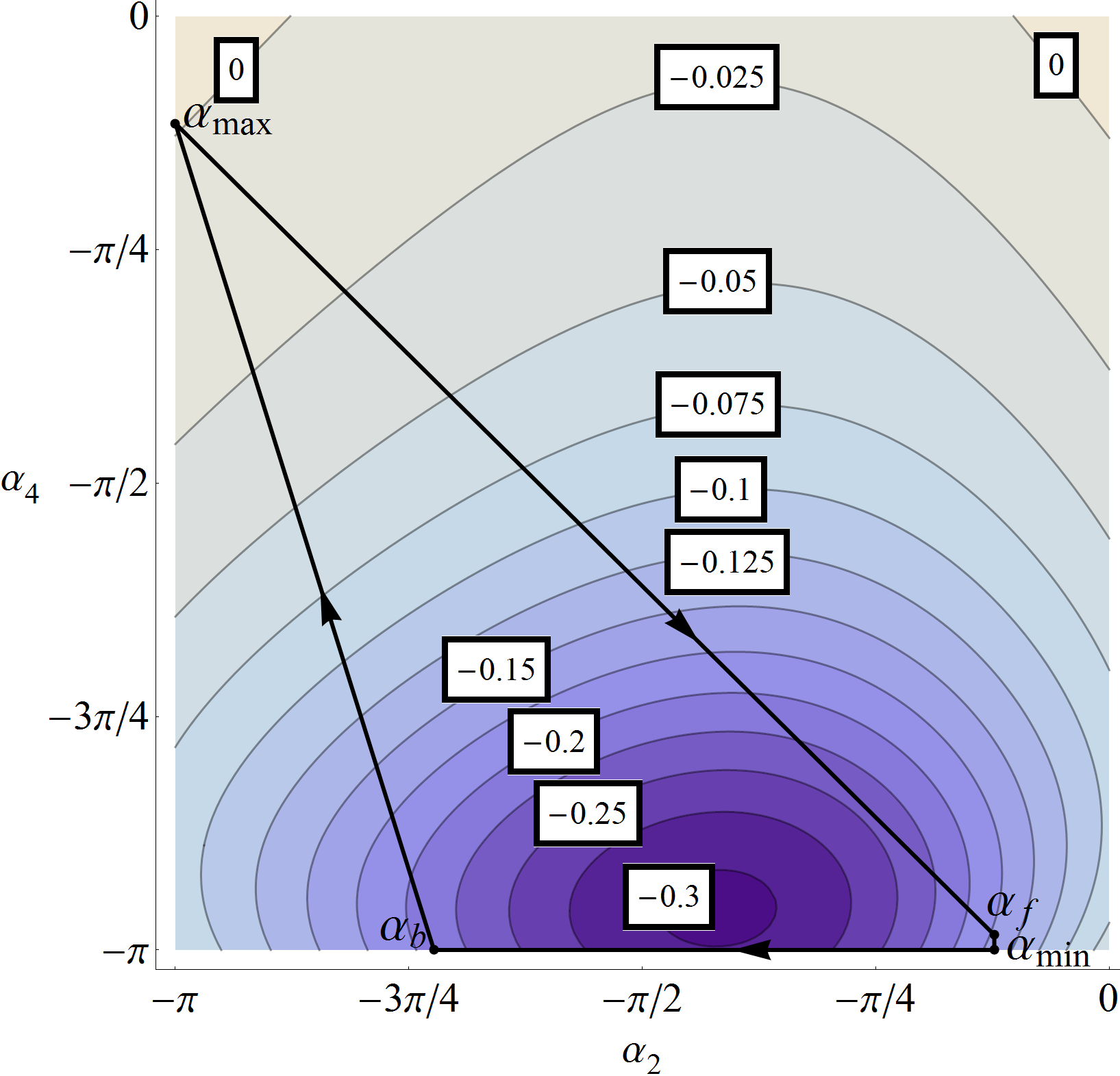}
\caption{Here $\{s_\mathit{in}, s_\mathit{out}\}=\{0.9818, 0.3158\}$, so the transition from $\bo{\alpha}_f= (-0.3867, -3.0910)$ to $\bo{\alpha}_\mathit{min}$ takes $0.0046$ seconds, and from $\bo{\alpha}_\mathit{min}$ to $\bo{\alpha}_b=(-2.2717, -\pi)$ takes $0.1711$ seconds. The loop is shown with constant $B(\bo{\alpha})$ contours, so the enclosed region gives an idea of the expected geometric phase contribution.}\label{fig:slowest}
\end{figure}

This idea is illustrated in Figure \ref{fig:slowest}, which has a slower leg movement when moving into pike, so that after a quarter second the arms are in place while the legs are not. This is indicated by the shape $\bo{\alpha}_f$, and the black vertical curve shows the additional leg movement required to reach $\bo{\alpha}_\mathit{min}$. When leaving pike position the arms move first to reach shape $\bo{\alpha}_b$, before both pairs of limbs move concurrently for a quarter second to complete the dive with shape $\bo{\alpha}_\mathit{max}$. 
In this part of the shape change the legs move at maximal speed.
While this results in rotational contribution from the geometric phase, the contribution from the dynamic phase is less due to the reduced time spent in pike position.

As the geometric phase given by \eqref{eq:green} involves integrating $B(\bo{\alpha})$ over the region enclosed by the loop $C$, having the absolute maximum of $B(\bo{\alpha})$ and its neighbouring large absolute values contained in this region will provide a more efficient (in terms of geometric phase per arc length) contribution towards the geometric phase. Maximising the overall rotation obtained therefore involves finding the balance between the dynamic and geometric phase contributions. We will now perform optimisation to determine the speed at which the arms and legs should move to achieve this.

\begin{figure}[p]
\centering
\hspace*{-10mm}\subfloat[$L=10$; $\{s_\mathit{in}, s_\mathit{out}\}=\{0.6093,0\}$]{\includegraphics[width=7.95cm]{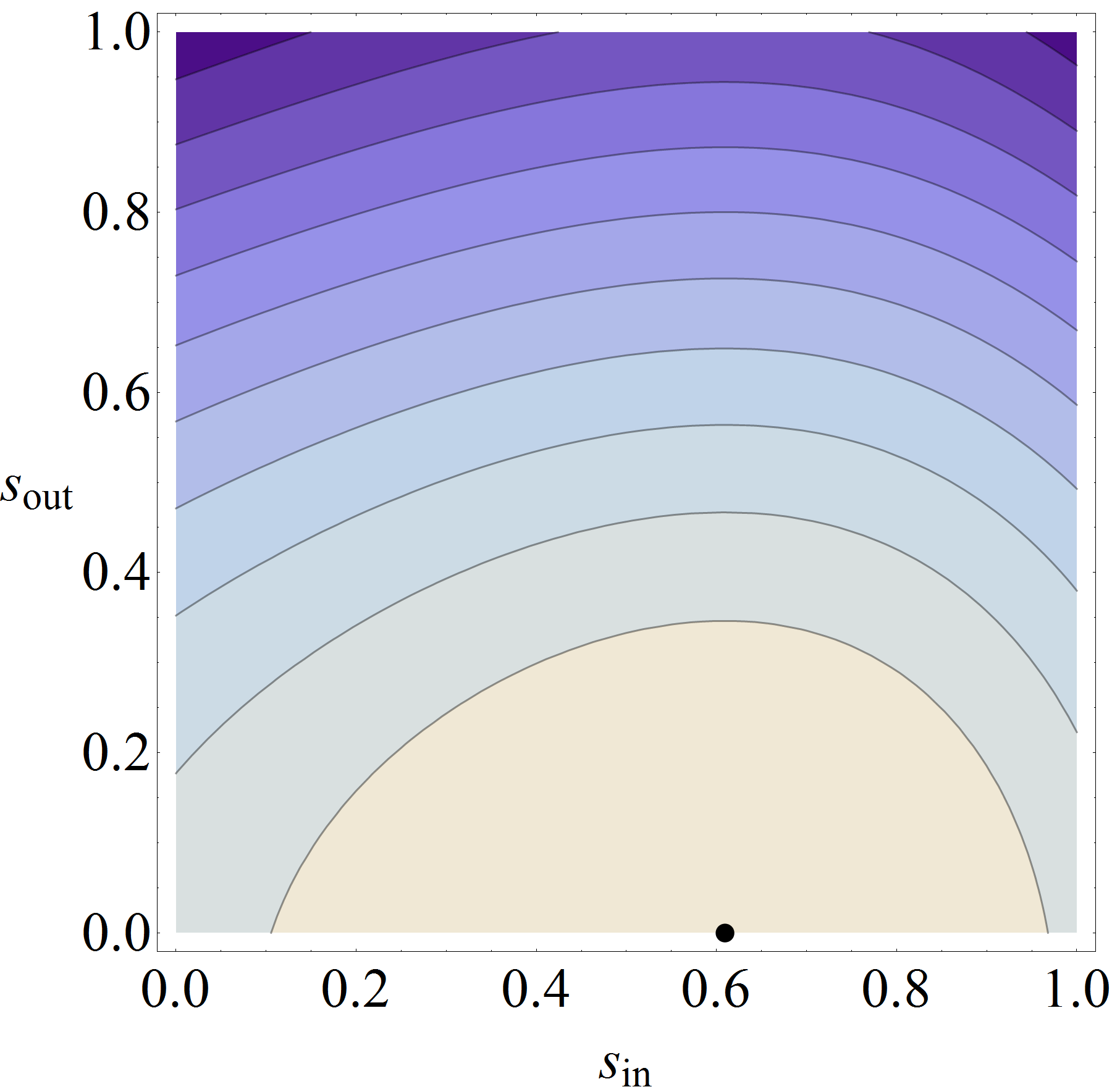}}
\hspace{2mm}\subfloat[$L=30$; $\{s_\mathit{in}, s_\mathit{out}\}=\{0.9818,0.3158\}$.]{\includegraphics[width=7.95cm]{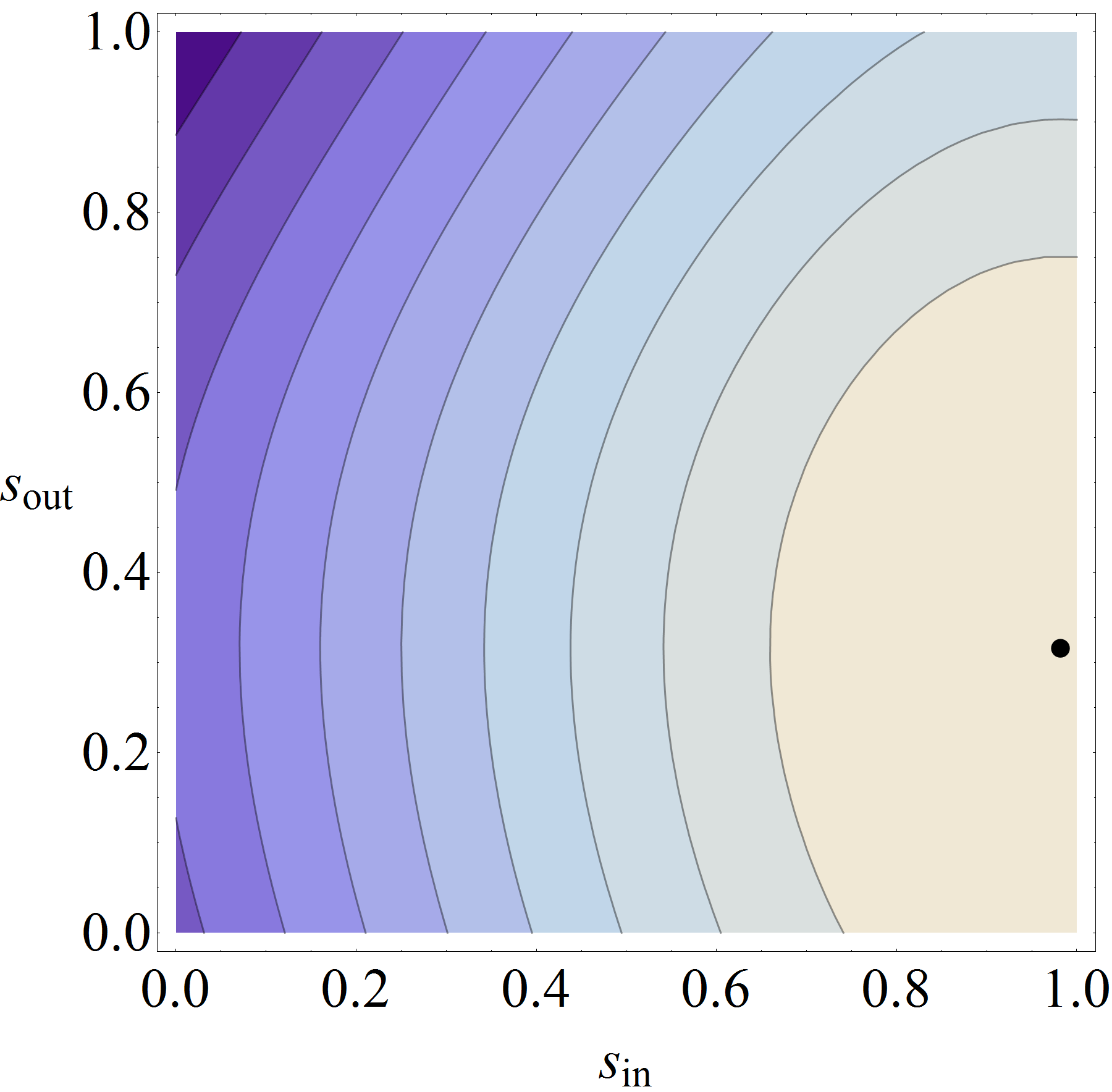}}\\
\hspace*{-10mm}\subfloat[$L=120$; $\{s_\mathit{in}, s_\mathit{out}\}=\{1,0.8593\}$.]{\includegraphics[width=7.95cm]{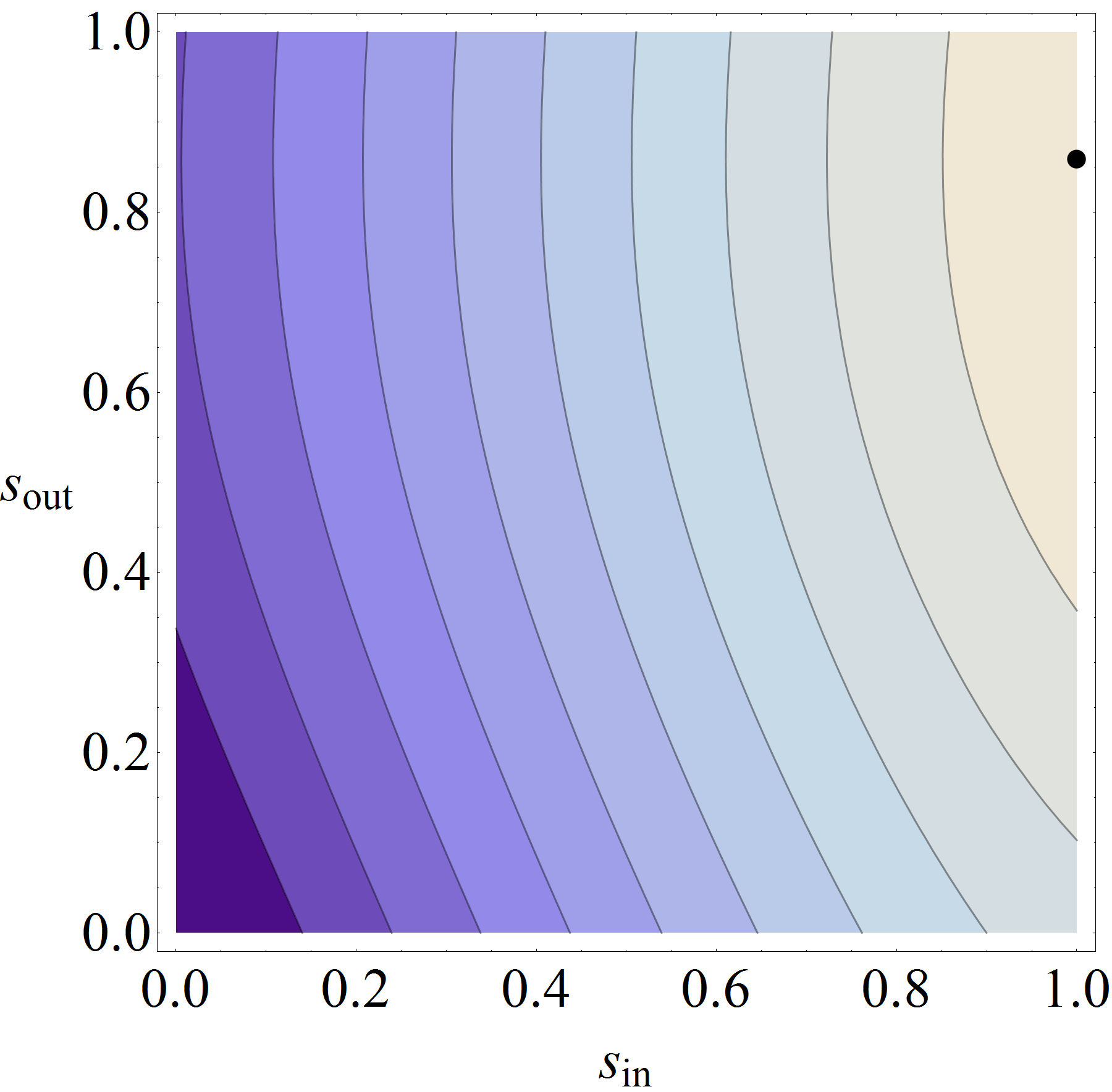}\label{subfig:s120}}
\hspace{2mm}\subfloat[$L=240$; $\{s_\mathit{in}, s_\mathit{out}\}=\{1,1\}$.]{\includegraphics[width=7.95cm]{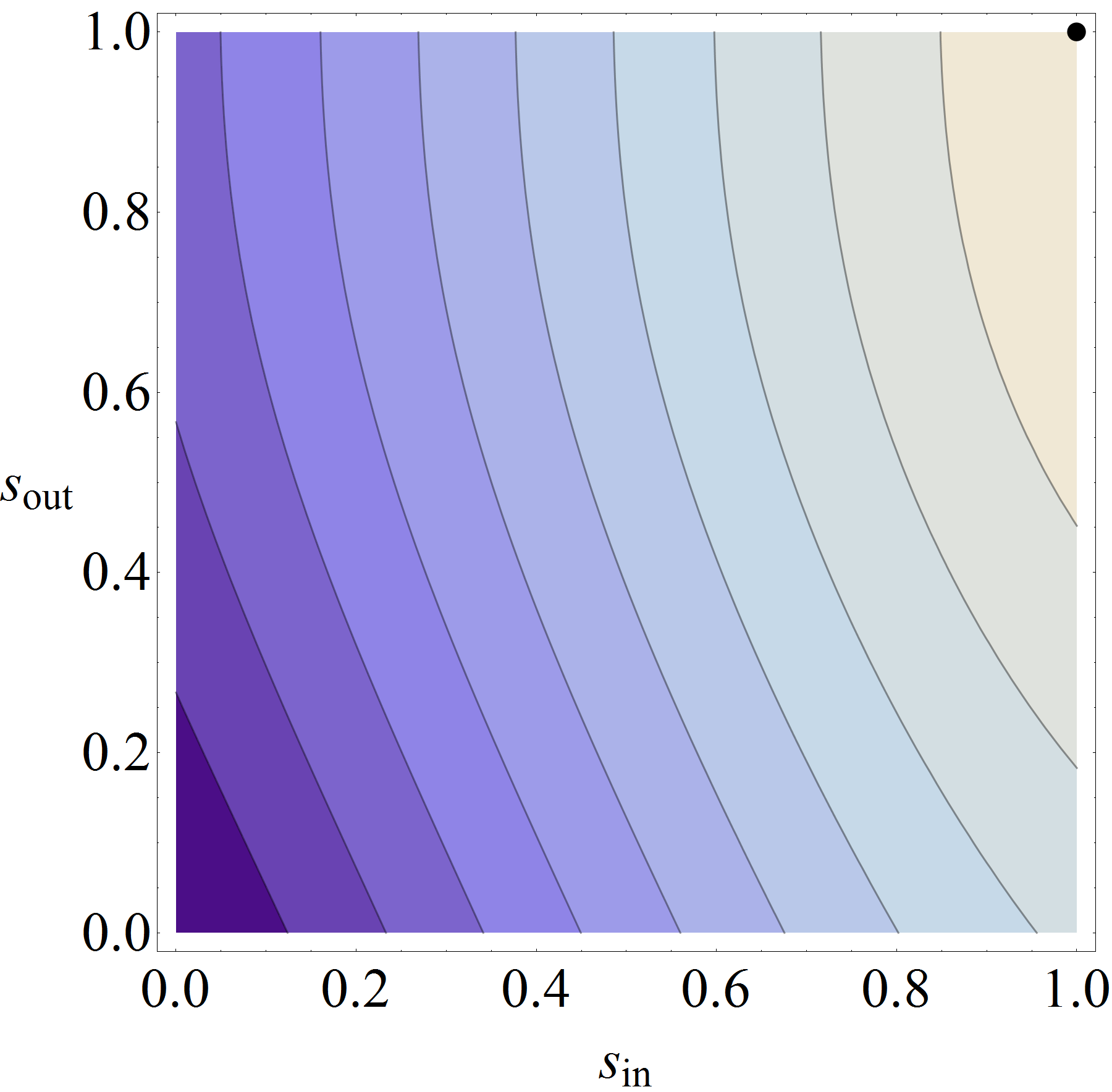}}\\
\caption{Contour plots of constant $\theta$ for four different $L$ values shown. The optimal $\{s_\mathit{in}, s_\mathit{out}\}$ in each case that maximises the overall rotation $\theta$ is specified and indicated by the black point.}\label{fig:s}
\end{figure}
\begin{figure}[t]
\centering
\includegraphics[width=12cm]{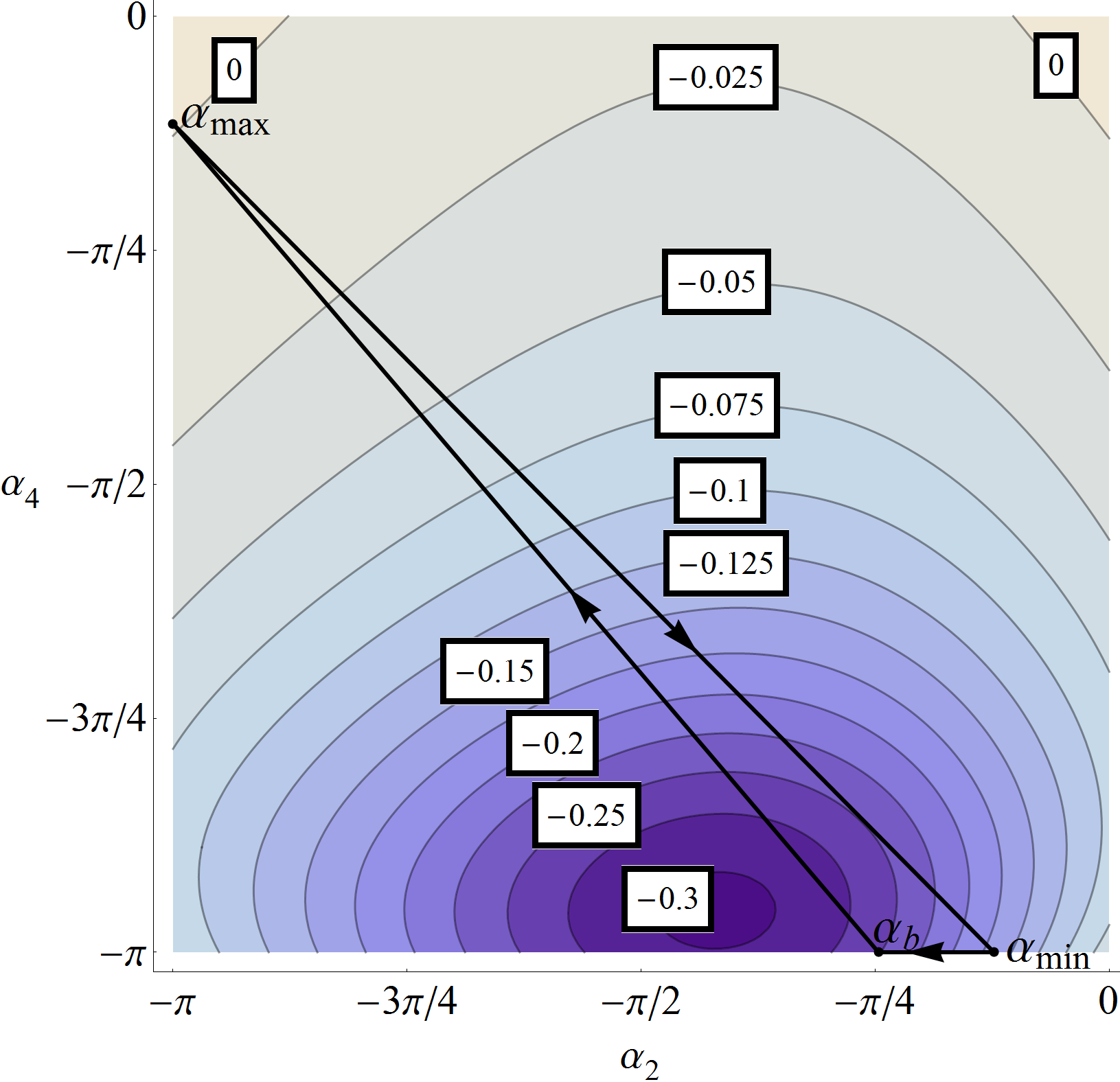}
\caption{Loop with $\{s_\mathit{in}, s_\mathit{out}\}=\{1, 0.8593\}$ shown with constant $B(\bo{\alpha})$ contours, and the point $\bo{\alpha}_b=(-0.7743, -\pi)$.}\label{fig:optimal}
\end{figure}
We define $s_\mathit{in}\in[0,1]$ to be the fraction of maximum speed for the leg movement into pike position, and $s_\mathit{out}\in[0,1]$ to be the fraction of maximum speed for the arm movement out of pike. These reduced speeds (for either arms or legs) will only be used when moving from $\bo{\alpha}_\mathit{max}$ to $\bo{\alpha}_f$ and $\bo{\alpha}_b$ to $\bo{\alpha}_\mathit{max}$, as the transitions from $\bo{\alpha}_f$ to $\bo{\alpha}_\mathit{min}$ and $\bo{\alpha}_\mathit{min}$ to $\bo{\alpha}_b$ will involve the appropriate limb moving at maximum speed to minimise the extra time spent in shape change. With this construction, the extra time required to move into and out of pike is $\tau_E(s_\mathit{in})$ and $\tau_E(s_\mathit{out})$, where
\begin{equation}
\tau_E(s) = (1-s)/4.\label{eq:extT}
\end{equation}
For the dive illustrated in Figure \ref{fig:fastest} we have $\{s_\mathit{in}, s_\mathit{out}\}=\{1,1\}$, thus both $\tau_E(s_\mathit{in}) = \tau_E(s_\mathit{out}) =0$. However, in general when $\tau_E(s_\mathit{in}) \neq 0$ and  $\tau_E(s_\mathit{out}) \neq 0$ the dive is composed of four pieces: the transitions from $\bo{\alpha}_\mathit{max} \rightarrow \bo{\alpha}_f$, $\bo{\alpha}_f \rightarrow \bo{\alpha}_\mathit{min}$, $\bo{\alpha}_\mathit{min} \rightarrow \bo{\alpha}_b$ and $\bo{\alpha}_b \rightarrow \bo{\alpha}_\mathit{max}$, which we denote as $\bo{\alpha}_i(t)$ for $i\in\{1,2,3,4\}$, respectively.
Let the transition time for each $\bo{\alpha}_i(t)$ be from zero to $\tau_i$, then $\tau_1=\tau_4=1/4$, $\tau_2=\tau_E(s_\mathit{in})$, $\tau_3=\tau_E(s_\mathit{out})$, giving the cumulative shape change time  
\begin{equation}
T_\Sigma = 1/2 + \tau_E(s_\mathit{in}) + \tau_E(s_\mathit{out}).
\end{equation} 
Combined with $T_\mathit{pike}$, which is the duration the athlete maintains pike position, we obtain the total airborne time
\begin{equation}
T_\mathit{air} = T_\Sigma + T_\mathit{pike}.\label{eq:Tair}
\end{equation}
Defining the shape change velocities in vector form to be
\begin{equation}
\bo{v}(s_\mathit{in},s_\mathit{out}) = 4\diag{(s_\mathit{out}, s_\mathit{in})}(\bo{\alpha}_\mathit{min}-\bo{\alpha}_\mathit{max}),
\end{equation}
the shape changing transitions can be written as 
\begin{align}
\bo{\alpha}_1(t) &= \bo{\alpha}_\mathit{max}+t\bo{v}(s_\mathit{in}, 1) & \bo{\alpha}_2(t) &= \bo{\alpha}_f+t\bo{v}(1,0) \nonumber\\
\bo{\alpha}_3(t) &= \bo{\alpha}_\mathit{min}-t\bo{v}(0,1) & \bo{\alpha}_4(t) &= \bo{\alpha}_b-t\bo{v}(1, s_\mathit{out}),\nonumber
\end{align}
where $\bo{\alpha}_f = \bo{\alpha}_1(\tau_1)$ and $\bo{\alpha}_b = \bo{\alpha}_3(\tau_3)$. 
Phases $2$ and $3$ disappear when $s_{in} = s_{out} = 1$ because the corresponding time spent in these phases is $\tau_E(1) = 0$.
Solving for the overall rotation obtained via \eqref{eq:theta} we get 
\begin{equation}
\theta(s_\mathit{in},s_\mathit{out})=\theta_\mathit{dyn}(s_\mathit{in},s_\mathit{out})+\theta_\mathit{geo}(s_\mathit{in},s_\mathit{out}),\label{eq:maxtheta}
\end{equation}
where the components are
\begin{align}
\theta_\mathit{dyn}(s_\mathit{in},s_\mathit{out}) &= L\left[\sum_{i=1}^4 \int_0^{\tau_i}I^{-1}\big(\bo{\alpha}_i(t;s_\mathit{in},s_\mathit{out})\big)\,\mathrm{d}t + I^{-1}(\bo{\alpha}_\mathit{min})T_\mathit{pike}\right]\nonumber\\
\theta_\mathit{geo}(s_\mathit{in},s_\mathit{out}) &= \sum_{i=1}^4 \int_0^{\tau_i}\bo{F}\big(\bo{\alpha}_i(t;s_\mathit{in},s_\mathit{out})\big)\cdot\bo{\dot{\alpha}}_i(t;s_\mathit{in},s_\mathit{out})\,\mathrm{d}t.\nonumber
\end{align}
The above components can be further simplified by substituting in \eqref{eq:Tair} to eliminate $T_\mathit{pike}$, and combining the shape change segments $\bo{\alpha}(t) = \bigcup_{i=1}^4 \bo{\alpha}_i(t)$ so that $t$ runs from zero to $T_\mathit{air}$ for the complete dive. This then gives
\begin{align}
\theta_\mathit{dyn} &= L\left[\int_0^{T_\mathit{air}}I^{-1}(\bo{\alpha})\,\mathrm{d}t -I^{-1}(\bo{\alpha}_\mathit{min})T_\Sigma + I^{-1}(\bo{\alpha}_\mathit{min})T_\mathit{air}\right]\label{eq:thetadyn}\\
\theta_\mathit{geo} &= \int_0^{T_\mathit{air}} F(\bo{\alpha})\cdot\dot{\bo{\alpha}}\,dt = \iint_A B(\bo{\alpha})\,d\tilde{A},
\end{align}
where the arguments $t, s_\mathit{in}$ and $s_\mathit{out}$ have been suppressed to avoid clutter. We see in \eqref{eq:thetadyn} that the constant $T_\mathit{air}$ provides an overall increase in $\theta_\mathit{dyn}$, but otherwise plays no role in the optimisation strategy. 
To obtain maximal rotation there is competition between having a large $B$ and a small $I$ for as long as possible. 
The angular momentum $L$ is an important parameter:
decreasing $L$ reduces the contribution from $\theta_\mathit{dyn}$ while $\theta_\mathit{geo}$ is unaffected, thus the geometric phase contribution has a greater impact on $\theta$ when $L$ is small. The essential competition in the optimisation in this model comes from the extra time $T_\Sigma$ taken to make the loop larger around large $B$ 
(which increases the geometric phase) and $T_\Sigma$ taken away from the time spent in pike position (which decreases the dynamic phase).
Our main result is that for a fixed total time $T_\mathit{air}$ and a moderate $L$ the total amount of somersault can be improved by using a shape change that has a different motion into pike than out of pike, and hence generates a geometric phase. The detailed results are as follows.

Figure \ref{fig:s} shows the optimal $\{s_\mathit{in}, s_\mathit{out}\}$ that maximises $\theta$ for different values of $L$. When $L=0$ this implies $\theta_\mathit{dyn}=0$, meaning the rotation is purely governed by $\theta_\mathit{geo}$. So by choosing $\{s_\mathit{in}, s_\mathit{out}\} = \{0,0\}$, the geometric phase is maximised and so too is the overall rotation. 
The case $L=0$ corresponds to the problem of the falling cat reorienting itself: rotation can be generated without angular momentum via a shape change. As $L$ increases the dynamic phase becomes proportionally larger, making it more important to enter pike position earlier, and hence $\{s_\mathit{in}, s_\mathit{out}\}\rightarrow\{1,1\}$ as $L$ gets larger. The limiting point $\{s_\mathit{in}, s_\mathit{out}\} = \{1,1\}$ occurs when $L=193.65$, where beyond this point maximum overall rotation is achieved by transitioning to pike position as fast as possible, as demonstrated in Figure \ref{fig:fastest}. For comparison, Figure \ref{fig:slowest} shows the optimal shape change trajectory when $L=30$.

When optimising $\theta$ in \eqref{eq:maxtheta} for a typical planar somersault with $L=120$, we see in Figure \ref{subfig:s120} that $\{s_\mathit{in}, s_\mathit{out}\}=\{1, 0.8593\}$ yields the maximal rotation, whose shape change trajectory is illustrated in Figure \ref{fig:optimal}. When compared to the dive with $\{s_\mathit{in}, s_\mathit{out}\}=\{1, 1\}$, the gain in overall rotation is $0.0189$ radians or $1.0836^\circ$. We see that in the optimal dive the limbs move at maximum speed into pike position, but when leaving the pike the arms move slower, thus providing a geometric phase contribution that exceeds the dynamic phase contribution lost by $1.0836^\circ$. This behaviour is due to the take-off and pike shapes being $\bo{\alpha}_\mathit{max}$ and $\bo{\alpha}_\mathit{min}$, but had we chosen different shapes then the observed result may differ. The dynamic phase benefits from being at (or close to) the minimum moment of inertia $I_\mathit{min}$ located at $\bo{\alpha}_\mathit{min}$, whereas the geometric phase favours enclosing a region with large magnitudes of the magnetic field $B$, where the absolute maximum is $0.3058$ occurring at $(-1.3148, -3.0043)$. 

Repeating the same computation with $L=30$ yields the optimal shape change trajectory shown in Figure \ref{fig:slowest}, and comparing this with the $\{s_\mathit{in}, s_\mathit{out}\}=\{1, 1\}$ dive yields an additional $0.2327$ radians or $13.3354^\circ$ in rotation, which is more significant than the $1.0836^\circ$ found for $L=120$. Although the addition of $1^\circ$ is irrelevant in practice, the model used is extremely simple and these results should only be considered as a proof of principle. We hope that by using more realistic models with asymmetric movements, the additional rotation obtained via geometric phase optimisation will yield a small but important contribution that transforms a failed dive into a successful one. The principles here are not limited to planar somersaults, but can also be utilised in other fields such as robotics and space aeronautics, 
where the benefit of the geometric phase will be large whenever the angular momentum is small.

\section*{Acknowledgement}
This research was supported in part by the Australian Research Council through the Linkage Grant LP100200245 “Bodies in Space” in collaboration with the New South Wales Institute of Sport.

\appendix
\section{Definition of $\bo{D}_i^j$, $\bo{\bar{D}}_i^j$ and $\bo{\tilde{D}}_i^j$}\label{app:D}
Consider a rooted tree where each $B_i$ is treated as a node with $B_1$ being the root (top most node). Let $B_{p(j)}$ denote the node who is the parent of $B_j$ in the tree, and $B_{c(j,i)}$ be the child of $B_j$, who is either $B_i$ or is the node with the direct line to $B_i$, i.e.\! an ancestor of $B_i$. An ancestor of $B_i$ is any node reachable by repeated proceedings from child to parent, e.g.\! the root of the tree $B_1$ is an ancestor to every other node. We can now give the general definition of the constant vectors $\bo{D}_j^i$ using the idea of trees. We have $\bo{D}_j^i=\bo{0}$ for $1\leq i\leq n$ and $1\leq j\leq n$, unless either $j=i$ and $j \neq 1$, or if $B_j$ is an ancestor of $B_i$ in the tree.
When this occurs we have
\begin{equation}
\bo{D}_j^i=\begin{cases}
-\bo{E}_j^{p(j)} & \mbox{if } j=i \mbox{ and } j \neq 1\\
\bo{E}_{j=\mathit{ref}}^{c(j,i)} & \mbox{if } j\neq i \mbox{ and } j = 1\\
-\bo{E}_j^{p(j)}+\bo{E}_j^{c(j,i)} & \mbox{if } j\neq i \mbox{ and } j \neq 1.
\end{cases}
\end{equation}
Next, we have  
\begin{align}
\bo{\bar{D}}_j&=\frac{1}{M}\sum_{i=1}^{n}m_i \bo{D}_j^i
\end{align}
where $M$ is the total mass, and this can be interpreted as the weighted mean of the $\bo{D}_j^i$'s. Finally, 
\begin{equation}
\bo{\tilde{D}}_i^j = \bo{D}_i^j-\bo{\bar{D}}_i
\end{equation}
is the difference between $\bo{D}_i^j$ and the weighted mean $\bo{\bar{D}}_j$.

\bibliographystyle{siam}
\bibliography{PlanarSomersault}
\end{document}